\documentclass[orivec]{llncs}

\usepackage[T1]{fontenc}
\usepackage{flafter}
\usepackage{graphicx}
\usepackage[caption=false]{subfig}
\usepackage{amsmath}
\usepackage{amsfonts}
\usepackage{amssymb}
\usepackage{amsxtra}
\usepackage{./mathSymbols}
\usepackage{xspace}
\usepackage{xargs}
\usepackage[firstpage,angle=0,fontsize=11pt]{draftwatermark}
\usepackage[dvipsnames]{xcolor}
\usepackage{enumitem}
\usepackage{algorithm}
\usepackage{algpseudocode}
\usepackage{listings}
\usepackage{orcidlink}
\usepackage{doi}
\usepackage{hyperref}
\usepackage{bookmark}

\bookmarksetup{
  depth=3,
}

\lstdefinelanguage{Modest}{
    morekeywords={
        if, else, then, palt, par, stop, tau, none, break, do, process, function, alt%
    },
    emph=[1]{
        const, transient%
    },
    emph=[2]{
        int, real, action, true, false, bool, option, datatype, array, property%
    },
    emph=[3]{
        A[], E<>, Pmax, Pmin%
    },
    morecomment=[l]{//},
    morecomment=[s]{/*}{*/},
    morestring=[b]{"},
    sensitive=true,
    escapeinside={(*@}{@*)}
}

\lstdefinestyle{ModestStyle}{%
  keywordstyle={\itshape},
  emphstyle=[1]{\itshape},
  emphstyle=[2]{\itshape},
  emphstyle=[3]{\underbar},
  commentstyle={\itshape},
}

\makeatletter
\lst@AddToHook{OnEmptyLine}{\vspace{-0.75\baselineskip}}
\makeatother

\hypersetup{
    colorlinks=true,
    linkcolor=black,
    filecolor=black,
    citecolor=black,
    urlcolor=black,
    pdftitle={Probabilistic Verification for Modular Network-on-Chip System}
}

\begin{document}

\title{%
    \texorpdfstring{%
        Probabilistic Verification for Modular Network-on-Chip Systems%
            \thanks{N.W. and Z.Z. gratefully acknowledge the support of the ECE Department at Utah State University. A.H. was supported by the EU's Horizon 2020 R\,\&\,I programme under MSCA grant 101008233 (MISSION), the Interreg North Sea project STORM\_SAFE, and NWO VIDI grant VI.Vidi.223.110 (TruSTy). S.R. and K.C. were supported in part by National Science Foundation (NSF) grant CNS-2106237. Any opinions, findings, and conclusions or recommendations expressed in this material are those of the author(s) and do not necessarily reflect the views of the NSF.}
        }%
        {Probabilistic Verification for Modular Network-on-Chip Systems}%
}

\author{%
    Nick Waddoups\inst{1}\smash{\raisebox{2.5pt}{\orcidlink{0000-0003-1938-8204}}}
    \and
    Jonah Boe\inst{2}\smash{\raisebox{2.5pt}{\orcidlink{0000-0002-6555-005X}}}
    \and
    Arnd Hartmanns\inst{3}\smash{\raisebox{2.5pt}{\orcidlink{0000-0003-3268-8674}}}
    \and
    Prabal Basu\inst{4}\smash{\raisebox{2.5pt}{\orcidlink{0000-0002-4860-1089}}}
    \and
    Sanghamitra Roy\inst{1}\smash{\raisebox{2.5pt}{\orcidlink{0000-0002-3927-1612}}}
    \and
    Koushik Chakraborty\inst{1}\smash{\raisebox{2.5pt}{\orcidlink{0000-0003-0228-2737}}}
    \and
    Zhen Zhang\inst{1}\smash{\raisebox{2.5pt}{\orcidlink{0000-0002-8269-9489}}}
}

\institute{%
    Utah State University, Logan, UT, USA \\
    \email{\{nick.waddoups, sanghamitra.roy, koushik.chakraborty, zhen.zhang\}@usu.edu}
    \and
    Hill Air Force Base, Davis County, UT, USA
    \and
    University of Twente, Enschede, The Netherlands \\
    \email{a.hartmanns@utwente.nl}
    \and
    Cadence Design Systems, San Jose, CA, USA
}

\maketitle

\SetWatermarkText{\raisebox{9.6cm}{%
    \hspace{10cm}%
        \href{https://doi.org/10.5281/zenodo.17247418}{%
        \includegraphics[%
            width=20mm,%
            keepaspectratio,%
            alt={VMCAI 2026 reuseable artifact badge}%
        ]{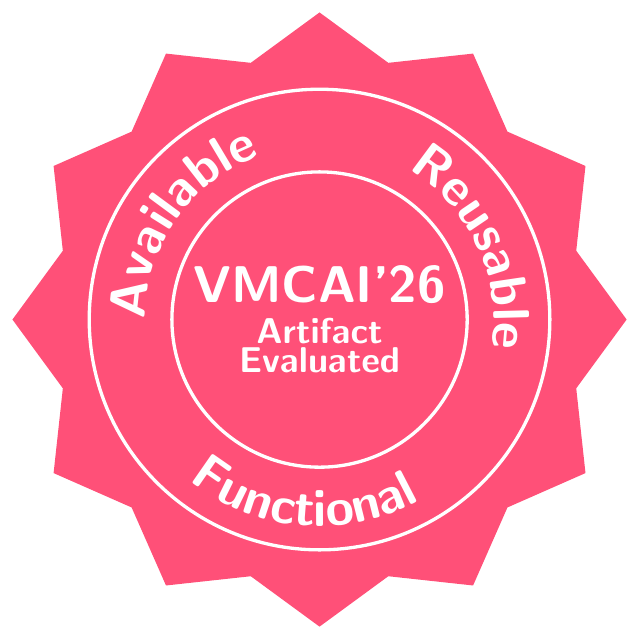}%
    }%
}}

\begin{abstract}

Quantitative verification can provide deep insights into reliable Network-On-Chip (NoC) designs. It is critical to understanding and mitigating operational issues caused by power supply noise (PSN) early in the design process: fluctuations in network traffic in modern NoC designs cause dramatic variations in power delivery across the network, leading to unreliability and errors in data transfers. Further complicating these challenges, NoC designs vary widely in size, usage, and implementation. This case study paper presents a principled, systematic, and modular NoC modeling approach using the \modest language that closely reflects the standard hierarchical design approach in digital systems. Using the \modestToolset, functional and quantitative correctness was established for several NoC models, all of which were instantiated from a generic modular router model. Specifically, this work verifies the functional correctness of a generic router, inter-router communication, and the entire NoC. Statistical model checking was used to verify PSN-related properties for NoCs of size up to \nxn{8}.

\end{abstract}

\section{Introduction}
\label{sec_introduction}

\noindent As modern digital systems grow in complexity, there is a need for efficient on-chip communications as the traditional shared data bus becomes overloaded when more sub-systems are added to a chip. \emph{Network-on-Chip} (NoC) designs are a widely-used alternative in computer chips that improve communication throughput and efficiency of a system by allowing more parallel communication.

NoCs have attracted substantial research attention for improving both reliability and performance, especially in complex designs such as System-on-Chips (e.g.,~\cite{Lecler2011,Tsai2012}). As modern semiconductor production integrates more subsystems on a single chip, from standard processing units to specialized hardware accelerators~\cite{Royannez2005}, NoCs are increasingly being adopted. However, their unique design challenges require correctness guarantees, particularly in safety-critical systems where faults can cause catastrophic failure~\cite{Naresh2017,Jiang2014}.

One major challenge in NoC designs is \textit{power supply noise} (PSN), which occurs with sudden shifts in traffic volume across the network. These shifts lead to fluctuations in the power drawn by each router and can result in data corruption or loss. Previous work~\cite{Basu2017} shows that PSN is magnified by the shift to smaller chip technology, with PSN increasing from $40\%$ of the supply voltage to around $80\%$ when moving an \nxn{8} NoC from 32\,nm to 14\,nm technologies.

Computer chips are designed in stages, starting with an architectural level specification, then moving onto an implementation in a hardware description language, and finishing with a physical layout before manufacturing. Design changes later in the cycle are significantly more costly than they would be if made earlier in the cycle. Incorporating probabilistic verification during the architectural stage of the NoC design flow allows for \textit{early} quantification of PSN in order to mitigate issues. While formal verification of NoCs is an active area of study, there is only limited work on probabilistic verification of small-scale NoCs~\cite{Roberts2021,Lewis2019}. We argue that probabilistic verification is \emph{required} to thoroughly check for PSN-related issues, as they are highly dependent on the inherent stochastic nature of network packet generation patterns.

Building on the second author's M.S. thesis work~\cite{boe2023probabilistic}, this paper has the following major contributions:

\begin{enumerate}[nolistsep,noitemsep,topsep=0pt,leftmargin=*]
    \item A highly modular NoC model, written in the probabilistic modeling language \modest~\cite{HahnHHK13,BohnenkampDHK06}, to allow easy construction of arbitrary-size NoCs and characterization of PSN (Section~\ref{sec:modular}). Previous work~\cite{Roberts2021} presents only probabilistic verification of a small-scale \nxn{2} NoC, modeled in a monolithic, non-modular manner, which does not scale. The parameterizable nature of our modular model enables conventional and probabilistic model checking of diverse NoC implementations.
    
    \item Verification of functional correctness of a generic router, inter-router communication, and full NoC model (Section~\ref{sec_verification}).

    \item Comparative model-checking: By reconfiguring our modular \nxn{2} NoC model into one equivalent to the earlier monolithic \nxn{2} NoC~\cite{Roberts2021}, we identified a deep-buried discrepancy in the modeling of packet consumption (Section~\ref{sec_results}). After resolving the discrepancies, our model successfully reproduced the results from~\cite{Roberts2021}, validating our 
    unique modular modeling approach.
    
    \item Probabilistic verification of PSN-related properties for NoCs of size up to \nxn{8}, demonstrating the scalability and flexibility of the modular approach, while uncovering PSN characteristics not captured in previous work~\cite{Roberts2021}.

    \item An analysis of the impacts of routing algorithms, demonstrating the frameworks ability to characterize PSN effects and reveal routing and scheduling strategies to improve reliability early in the design process.
    
\end{enumerate}

\section{Related Work}
\label{sec_related}

\paragraph{Probabilistic Modeling Checking.} 

Research and development of probabilistic verification tools has been carried out for decades, stemming from research in the 1980s~\cite{Katoen2016}. Modern research has produced probabilistic verification tools such as \modest~\cite{Hartmanns2014}, \prism~\cite{Kwiatkowska2011}, \storm~\cite{Hensel2022}, among others~\cite{Jeppson2023,Roberts2022,Hahn2014,Song2012}. They have been widely applied to fields from flight controllers~\cite{Wang2017} to hardware security and reliability~\cite{Mundhenk2015,Fang2014,Maes2013,Kumar2010} to synthetic biology~\cite{Taylor2023,Buecherl2021,Madsen2012,Lakin2012}. The \modestToolset can analyze models with quantitative data~\cite{Hartmanns2014} and includes a built-in statistical model checker~\cite{Budde2018} used to generate PSN probabilities in this work.

\paragraph{NoC Verification.}

Previous NoC verification work mainly focuses on functional correctness of NoCs. It has been used to verify the correct routing behavior~\cite{Taylor2022,Salamat2016}, check security properties~\cite{Sepulveda2018}, and evaluate performance~\cite{Alhubail2019}. These works have led to a \emph{qualitative} understanding of NoCs, but cannot fully quantify critical properties involving reliability. 

\paragraph{Probabilistic NoCs in Modest.}

Previous probabilistic verification for NoCs~\cite{Roberts2021,Lewis2019} was performed using \modest. \cite{Roberts2021} developed a \nxn{2} NoC to quantify the effects of PSN using probabilistic model checking. This model was developed as a proof of concept, but was modeled in a \textit{monolithic} fashion where all behavior and state variables were explicitly declared. This style of modeling is not scalable, and mistakes are easy to make. The goal of this work is to significantly enhance the scalability and usability of the concepts in~\cite{Roberts2021} to characterize PSN in NoCs.

\section{Preliminaries}
\label{sec_preliminaries}

\subsection{NoC Architecture}

\begin{figure}[hbt]
    \centering
    \subfloat[]{%
        \centering
        \includegraphics[alt={Diagram of a 2 by 2 Network-on-Chip}]{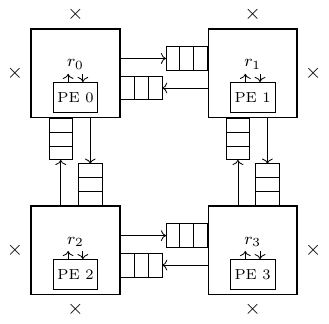}%
        \label{fig:2x2config}
    }%
    \subfloat[]{
        \centering
        \includegraphics[alt={Detailed diagram of a single router}]{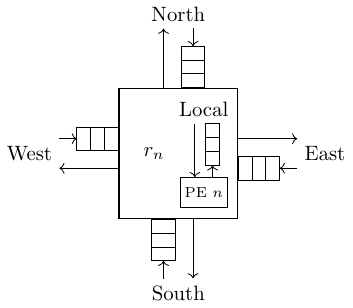}%
        \label{fig:Router}
    }%
    \caption{A \nxn{2} NoC architecture and an individual router}
    \label{fig:exRouterAndNoC}
\end{figure}

\noindent The NoC architecture presented in this paper is a variable-size square mesh network composed of individual routers. Figure~\ref{fig:2x2config} shows a \nxn{2} NoC composed of four individual routers and Figure~\ref{fig:Router} shows an individual router.

A \nxn{2} NoC is made up of four routers in a grid pattern. Each router is responsible for receiving packets and routing them to their destination. Routers are denoted using \ri{i}, where $i$ is the router ID.

A router contains first-in-first-out (FIFO) fixed-size \textit{buffers} to store incoming network packets and \textit{channels} to send packets. Buffers are represented by small blocks (``\(\Box\!\Box\!\Box\)'') and channels by arrows (``\(\rightarrow\)'') in Figure~\ref{fig:exRouterAndNoC}. Additionally, ``\(\times\)'' indicates that the buffer and channel on that side of the router are not connected which occurs when a router is positioned at the edge of the network.

A specific buffer in a router is addressed as \rib{i}{b}, where \(b\) is the buffer label and \(i\) is the router ID. The buffer is described either by its position relative to a router, such as \rib{i}{\mathit{North}}, or generally, such as \rib{i}{\mathit{Source}}. In digital design, a fixed-size FIFO buffer has queue-like semantics where elements are inserted in the back and popped from the front. In our models, a channel represents physical wires that carry network data from one router to another.

Packets are used to communicate data between routers in a NoC, similar to packets in a traditional computer network. Starting at one router, a packet is moved from router to router through the network until it reaches its destination. Packets are introduced into and removed from the NoC by a \textit{processing element} (PE) connected to each router via the local buffer and channel. A PE could be a CPU, accelerator core, memory cache, or other digital circuit. To move across the network, packets are (1) popped from the front of a buffer by a router, (2) routed through a channel towards their destination, and (3) pushed into the buffer connected to the channel.

As a digital circuit, NoCs are synchronized by a global clock. In our model, each channel can only be used once per clock cycle to send a packet. Additionally, in our model, each buffer can only accept one new packet per clock cycle. The sending of packets happens simultaneously (i.e., during the same clock cycle) between all NoC routers.

In digital design, a packet is often split up into smaller components known as \textit{flits}~\cite{Basu2017} due to limitations of channel bit-width. When a packet is transferred from one router to another, it is broken up into flits that are transmitted contiguously and sequentially over a channel, with one flit transferred per clock cycle. The number of flits in a packet is determined by the bit width of the channel, the size of the packet in bits, and the protocol being used.

The NoC model in this paper assumes a \emph{single-flit packet} that can be transferred between routers in one clock cycle for the following reasons:

\begin{enumerate}
    \item This work is targeted at characterizing PSN \textit{early} in the design cycle, before the implementation of the NoC is complete. As such, the channel width and packet size may not be known when this modeling is intended to take place.

    \item This work focuses on characterizing PSN based on high-level traffic patterns, routing algorithms, and NoC size. As such, our main consideration is to look at the high-level of packet traffic, not the individual movements of flits in the system. The abstraction presented in this work forgoes verification of circuit specific details for scalability and flexibility, which is desirable early in the design cycle. For more in-depth methods that look at characterizing PSN later in the design cycle, see~\cite{Basu2017}.

    \item Using single-flit packets greatly improves the tractability of the state space for verification. Roughly, a single-flit packet model has a state space complexity of \(O(S^n)\), where \(S\) is the number of states for each buffer, and \(n\) is the number of buffers. Modeling multi-flit packets introduces many other possible states for each buffer, which would worsen the state space explosion problem and render verification intractable for the model.
\end{enumerate}

Because we assume a single-flit packet, \textit{packets} and \textit{flits} are equivalent in our model as they each carry only a small piece of data and can be transferred from router to router in one clock cycle.

\subsection{Routing Algorithm}\label{sec:routingAlg}

Similar to computer networking, the path a flit takes to get from the sender to the receiver depends on the implemented routing protocol. The NoC design presented in this work allows \textit{any} NoC routing algorithm to be modeled. 

Results shown in this paper use the \textit{X-Y routing algorithm}: It moves flits first horizontally across the NoC until they are in the same column as the destination router, and then moves them vertically until they reach their destination. This routing algorithm was chosen because it is deterministic, deadlock free, and is the same algorithm used in~\cite{Roberts2021}. Additionally, the modularity of this work provides a framework for implementing other routing algorithms. More details about our implementation of the X-Y routing algorithm are given in Appendix~\ref{app:sec:processes:advance_router}.

\subsection{Router Arbitration}\label{sec:prelim:router_arb}

An important aspect of the router is the order in which each router's buffers are \textit{serviced}, i.e. when a flit is removed from a buffer and sent through a channel to a destination. This order determines a router's fairness and is also critical to ensuring correct operation. An incorrect algorithm may deadlock, starve a particular router, or allow multiple flits to be routed per clock cycle across a single channel. We discuss routing correctness in Section~\ref{sec:modular}.

Our NoC modeling framework uses CTL formulas of the form "forall globally" (\ensuremath{\forall\Box}) and "exists eventually" ($\exists\Diamond$) to ensure the functional correctness of a single router model and the composition to form an \nxn{n} NoC. \emph{Probabilistic CTL} (PCTL)~\cite{Hansson1994} is used to quantitatively check PSN-related properties. The PCTL syntax relevant to this work is 
\ensuremath{\mathbf{P}_{\sim q}(\Diamond^{\leqslant K}\Psi)}. 
That is, whether the probability that the state formula $\Psi$ is eventually true within $K$ steps lies within the interval specified by \ensuremath{\sim\!\! q}, where \ensuremath{\sim\; \in \{<,\leqslant,\geqslant,>\}} and $q \in [0,1]$. Additionally, the probabilistic operator $\mathbf{P}_{\sim q}$ can be replaced by $\mathbf{P}_{=?}$, indicating that the actual probability of the path formula $\Diamond^{\leq K}\Psi$ is of interest. For a complete definition of PCTL's syntax and semantics, see~\cite{Kwiatkowska2007}.

\subsection{Power Supply Noise (PSN)}
\label{sec:prelim:psn}

PSN has two components in a NoC: \textit{resistive noise} -- the product of the inherent resistance $R$ of a circuit and the current $I$ drawn, and \textit{inductive noise} -- the rate of change of current $\frac{\Delta I}{\Delta t}$ through the inherent inductance $L$ of the circuit. The voltage drop due to PSN can radically degrade the delay of various on-chip circuit components, causing \textit{timing errors}\footnote{A timing error is a fault caused when a pipe stage delay exceeds the clock period, leading to an erroneous value appearing in the network. In practice, this would appear as corrupted, dropped, or spurious data.} in the network. Existing approaches to mitigate PSN are a far cry from a truly reliable NoC design paradigm, due to the lack of worst-case peak PSN guarantees~\cite{Basu2017,Rajesh2015}.

Instead of directly measuring $I$, the model measures the activity of the router. Router activity is determined by the number of packets a router moves during a single clock cycle, ranging from zero (no packets) to five (all buffers used). This abstraction is supported by~\cite{Basu2017}, which shows that high activity correlates to higher current, and an abrupt change in router activity leads to a high rate of current change. Like~\cite{Roberts2021,Lewis2019}, PSN is tracked over clock cycles by introducing two counters, \resist and \induct. These counters represent the number of clock cycles where the router activity is over a user-specified router activity threshold $T$, where $T \in [0,5]$.

Property~\ref{eq:res} below is used to check the probability that the \resist counter will eventually exceed some threshold $K$ within $N$ clock cycles. Property~\ref{eq:ind} shows a similar property for determining the \induct probability within $N$ clock cycles. Note that clock cycle progress is a \emph{reward annotation} to certain transitions, i.e., \ensuremath{\mathrm{accumulate}(\mathit{clk})}, instead of being encoded in the structure of an expanded state space.

\begin{equation}
    \mathbf{P}_{=?}(\Diamond^{[\mathrm{accumulate}(\mathit{clk}) \leqslant N]}\: {\resist} \geqslant K)
    \label{eq:res}
\end{equation}

\begin{equation}
    \mathbf{P}_{=?}(\Diamond^{[\mathrm{accumulate}(\mathit{clk}) \leqslant N]}\: {\induct} \geqslant K)
    \label{eq:ind}
\end{equation}

This work characterizes PSN using the same method as~\cite{Lewis2019,Roberts2021}. This method abstracts away circuit-specific details in order to characterize PSN at a behavioral level. This is done because (1) early in the design process, engineers are often more concerned with high-level behavior, (2) circuit-level netlist models are not available early in the design process, and (3) once made available, circuit-level netlist models are intractable for formal verification. This abstraction to a behavioral characterization of PSN in NoCs cannot be directly converted or compared to measured PSN on a digital circuit. However, the correlation between router activity and PSN shown in~\cite{Basu2017} indicates that the high-level abstraction used in this work enables valid characterization of PSN early in the design stage, providing key insights into the expected profile of PSN in the design. 

\subsection{Statistical Model Checking (SMC)}

SMC is a quantitative verification approach that enables verifying the probability that a system model satisfies some property of interest. SMC uses data collected from random simulation runs to approximate the probabilistic properties of a model~\cite{Legay2010}. As a result, the confidence level of these approximations depends on the number of runs executed, with more executions resulting in higher confidence. Models containing rare events may take hundreds of thousands of runs to achieve accurate results. This work uses the SMC tool \lstinline{modes}~\cite{Budde2018}, part of the \modestToolset~\cite{Hartmanns2014}. While the \modestToolset also contains tools for probabilistic model checking~\cite{BaierAFK18} (which uses exhaustive state space exploration), this work primarily uses SMC for its unparalleled scalability.

\subsection{The Modest Toolset}

The presented NoC architecture, routing algorithms, and properties are modeled and specified using the \modest formal modeling language. Properties are then checked with the \modestToolset. While a comprehensive description of \modest is outside the scope of this work, some concepts critical to the function of the model are outlined here. The \lstinline{process} construct in \modest models an asynchronous process and may contain local variables, atomic assignments, and synchronizing actions. Processes can be composed using either \textit{sequential} or \textit{parallel} composition. Parallel processes are synchronized through \textit{actions}, which act as a synchronization barrier in a parallel composition where all participating processes must synchronize before performing the next step. A short example of \lstinline{process} and composition syntax is given in Code Segment \ref{cs:def:par_seq_comp}. We refer the interested reader to~\cite{hartmanModestTutorial} for a tutorial and reference for \modest.

The semantics of \modest consists of two components: (1) a \emph{symbolic semantics}~\cite[Sect.~4.2]{BohnenkampDHK06} that maps syntax to a symbolic \emph{stochastic timed automaton}~(STA) and (2) a \emph{concrete semantics}~\cite[Sect.~5]{BohnenkampDHK06} that transforms the STA into a \emph{timed probabilistic transition system}~(TPTS), a form of uncountable-state Markov decision process. In our work, we use a subset of \modest where the concrete semantics yields a \emph{discrete-time Markov chain}~(DTMC).

\subsection{Rationale for Choosing \modest}

\modest is our preferred modeling language and toolset for this work, as it is the \textit{only} mature probabilistic formal modeling language that supports flexible data structures for modular modeling, such as arrays, linked lists, and structs. These data structures are \emph{not} supported by the other mature probabilistic modeling language, that of \prism~\cite{Kwiatkowska2011}, yet they are key to our goal of developing a \textit{highly modular} NoC model.

While the NoC model analyzed in this work is synchronous, \modest and \prism are inherently asynchronous. To resolve this, we can use the synchronizing actions offered by both \modest and \prism that force specified parts of the model to be synchronous. Other naturally synchronous formal modeling languages such as nuXmv~\cite{cavada_nuxmv_2014} or Lustre~\cite{caspi_lustre_1987,champion_kind2_2016} lack the required support for specifying models with probabilistic characteristics.

Finally, the \modestToolset has been actively maintained for many years and provides a suite of tools for statistical and probabilistic model checking of PCTL properties, with recently added support for checking CTL properties.

\begin{lstlisting}[
    float=htbp,
    caption={Process and Parallel Composition Syntax in \modest},
    label={cs:def:par_seq_comp},
    captionpos=b,
    language=Modest,
    style=ModestStyle
]
process P0() { /*...*/ } // process declaration
par { :: P0() :: P1() } // parallel composition
P0(); P1(); // sequential composition
\end{lstlisting}
\section{Modular Design of the Formal NoC Model}
\label{sec:modular}

\noindent We provide a framework for creating arbitrary \nxn{n} two-dimensional mesh NoCs in the \modest language. The NoC model is a composition of instances of a router process, modeled using the \lstinline{process} construct in \modest. This modular style was influenced by modern digital design using \emph{hardware description languages} such as Verilog \cite{system_verilog_lrm} or VHDL \cite{vhdl_lrm}, where modular design is essential for scalable designs and commonly used across all hardware designs. A modular design is defined by the clean separation of circuitry between different modules, where each module has a well defined interface that can be connected to other modules. Our modular design of the NoC model provides scalable results beyond that of previous works and more closely aligns with digital design practices.

This work presents a NoC model with easy customization of design parameters, including topological size, flit injection pattern, buffer size, and routing algorithm. The PSN characterization and verification of functional correctness is specified as properties \emph{independent} of the specific model allowing for comparison of PSN characterization between models. Parameters such as topological size, buffer size, and PSN thresholds can be easily updated without knowledge of the \modest language. Customization to behavior (e.g. packet injection pattern, routing algorithm, or priority arbitration) require writing \modest code.

The full specification of the NoC model is too complex to fit within the page limit of this work. For this purpose, Appendix~\ref{app:sec} has been made available with more details, examples, and results that support this work. 

\subsection{NoC Construction using Parallel Composition}
\label{sec:modular:composition}

An \nxn{n} NoC is described in our framework as the parallel composition of $n^2$ router processes and a clock process. For example, the \nxn{2} NoC shown in Figure~\ref{fig:2x2config} is a parallel composition of four routers and a clock, i.e., \ensuremath{\ri{0} \parallel \ri{1} \parallel \ri{2} \parallel \ri{3} \parallel Clock}. Code Segment~\ref{cs:Par} shows the parallel composition in \modest, where each router is given its unique ID as input parameter with $\mathrm{ID} \in [0,n^2-1]$. The router process demonstrates how a modular design improves the scalability of formal NoC models, as constructing a NoC of arbitrary size is accomplished by instantiating more routers.

\begin{lstlisting}[
    float=htbp,
    caption={Creation of a \nxn{2} NoC in \modest},
    label={cs:Par},
    captionpos=b,
    language=Modest,
    style=ModestStyle
]
par { :: Clock() :: Router(0) :: Router(1) :: Router(2) :: Router(3)}
\end{lstlisting}

\subsection{Router Process}
\label{sec:modular:router}

The \lstinline{Router} process models the generic router model in Figure~\ref{fig:Router}. This router has four input buffers and four output channels for interfacing with neighboring routers and a local input buffer for introducing new packets. The local buffer is the only way new packets are introduced into the NoC. The router model can be instantiated many times with a unique ID (Code Segment~\ref{cs:Par}).

Each router maintains variables that define the state of the NoC, such as the elements in each buffer, previously serviced buffers, and various boolean flags that indicate what actions a buffer can take. These variables are stored in a global array of router objects. When instantiating a router process, the given \lstinline{id} acts as an index into the array of router objects. These router objects cannot be completely default initialized due to design choices of \modest, but a script is provided to automate this process as described in the Appendix~\ref{app:sec:python}.

The buffers are modeled as a functional programming language-style \textit{list}. This implementation efficiently models a FIFO queue, and allows for buffer size to be easily specified by a single global constant.

As shown in Code Segment~\ref{cs:proc:Router}, the router process is divided into five major sub-processes: generating new flits for the local buffer, preparing the router for flit advancement, advancing flits to their respective output channels, updating the buffer scheduling priority, and updating the PSN variables. 
After the \lstinline{nextClockCycle} action, the router process calls itself recursively, which models the repeated execution of the same five sub-processes in every clock cycle.

Although the modular style allows for scalability, code reuse, and customization, it also initially had synchronization issues between routers that were not present in the previous work~\cite{Roberts2021} because \cite{Roberts2021} did not use parallel composition to model the NoC. The parallel composition shown in Code Segment~\ref{cs:Par} is asynchronous, which means that each router instance executes its actions at its own speed without waiting for other routers.

\begin{lstlisting}[
    float=htbp,
    aboveskip=2ex,
    caption={\lstinline{Router} Process in \modest},
    label={cs:proc:Router},
    captionpos=b,
    language=Modest,
    style=ModestStyle
]
process Router(int id) {
  GenerateFlits(id); PrepRouter(id);
  AdvanceRouter(id); UpdatePiority(id);
  nextClockCycle; /* synchronizing clock action */
  Router(id) /* recursive call models next clock cycle */ }
\end{lstlisting}

Two key issues presented themselves when initial verification attempts were performed on the NoC. First, the asynchronous composition meant that there was no guarantee that all routers would be synchronized. Because of this, one router could get ahead of its neighbors, effectively allowing multiple steps to be taken per clock cycle, which is not possible in a synchronous digital system. This error was fixed in~\cite{boe2023probabilistic} by introducing a global synchronization action, similar to \lstinline{nextClockCycle} in Code Segment~\ref{cs:proc:Router}.

In \modest, synchronization actions act as a barrier that every process must reach before proceeding. When the \lstinline{nextClockCycle} action is reached, every process must wait for every other process to also reach the \lstinline{nextClockCycle} action before proceeding. This prevents routers from performing multiple steps per clock cycle, and ensures that per-cycle PSN results are correct.

Unfortunately, a single \lstinline{nextClockCycle} action is not sufficient to prevent all synchronization errors in the router. While the \lstinline{nextClockCycle} action synchronizes the \lstinline{Clock} and \lstinline{Router} processes, errors can still occur where one flit passes through multiple routers in one clock cycle. This is a write-before-read conflict, where one router may call \lstinline{AdvanceRouter} before another has called \lstinline{PrepRouter}, and is not a valid NoC behavior.

For example, if $\ri{0}$ has a flit destined for $\ri{3}$ in the \nxn{2} NoC depicted in Figure~\ref{fig:2x2config}, it should take a minimum of two clock cycles to reach its destination. On clock cycle 1, the flit is sent from $\ri{0}$ to $\ri{1}$. Then, on clock cycle 2, it is sent from $\ri{1}$ to $\ri{3}$. However, if $\ri{0}$ executes \lstinline{AdvanceRouter} before $\ri{1}$ executes \lstinline{PrepRouter}, the flit could travel from $\ri{0}$ to $\ri{1}$ to $\ri{3}$ in a single clock cycle.

A second synchronization action \lstinline{sync} was inserted between the \lstinline{PrepRouter} and \lstinline{AdvanceRouter} processes in \cite{boe2023probabilistic} and this successfully prevented write-before-read conflicts. However, when performing functional verification on the resulting model, the state space would explode due to the need to store states resulting from interleavings of each router \emph{independently} executing each sub-process. These interleavings were not desired  as they would not occur in a synchronous digital system. In this work, we inserted \textit{fine-grained synchronization actions} into every sub-process so that all routers in the parallel composition executed in lock step. This model update not only alleviates state space explosion, but also more faithfully models synchronous digital hardware, where assignments across all components execute in lock step during every clock cycle.

\subsection{Synchronizing Actions}
\label{sec:modular:sync_actions}

The \modest language is inherently asynchronous while most digital systems are synchronous with a global clock synchronizing all the updates in lock-step. To ensure that our \modest model correctly models a synchronous system, we carefully employ fine-grained synchronization actions among all processes. In \modest, each assignment block -- assignments within \lstinline|{= =}| braces -- executes atomically. As described in this section, the NoC model is broken up into many processes, where each process handles one element of the model. Parallel composition of these processes produces one single model that acts as a NoC. In parallel assignment in \modest, each unique interleaving of assignment blocks is explored. This produces many interleavings of actions, which is useful for many problems that \modest can model. However, when modeling synchronous digital designs, we are not concerned with interleaving states, we are only concerned with the changes that occur with the synchronous clock.

\begin{figure}[hbt]
    \centering%
    \subfloat[][{Without actions}]{
        \centering%
        \includegraphics[alt={State space without actions-4 total states}]{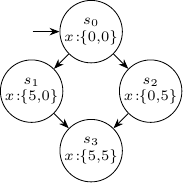}%
        \label{subfig:interleaved_ss}%
    }%
    \hspace*{2em}%
    \subfloat[][{Digial semantics}]{
        \centering%
        \raisebox{3.5ex}{\includegraphics[alt={Synchronous digital updates}]{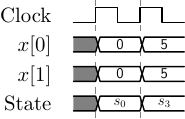}}%
        \label{subfig:digital_update}%
    }%
    \hspace*{2em}%
    \subfloat[][{With actions}]{
        \centering%
        \raisebox{1ex}{%
            \hspace{1em}%
            \includegraphics [alt={State space with actions-2 total states}]{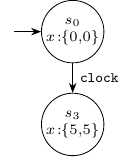}%
            \hspace{1em}%
        }%
        \label{subfig:action_ss}%
    }%
    \caption{Adding Actions to Reduce Unnecessary State Space}
    \label{fig:action_reduce_ss}
\end{figure}

Code segment~\ref{cs:interleaving_example} shows an example of a process that updates an element in array \lstinline{x} based on the input \lstinline{id}. The possible states including interleavings for these assignments are shown in Figure~\ref{subfig:interleaved_ss}. If this were modeling a digital circuit, and \lstinline{x} was a digital register block that updated with the clock, states \(s_1\) and \(s_2\) would never be observed due to the synchronous semantics of the digital system, as shown in the clock diagram in Figure~\ref{subfig:digital_update}. In practice, this problem of unnecessary interleaving causes an exponential explosion of state space that makes the modular model challenging to verify with the available hardware. This issue was not a problem for previous work~\cite{Roberts2021} because that model was not made of modular processes composed together, but was rather represented by a large process that orchestrated each part of the model and updates occurred in the same assignment block.

\begin{lstlisting}[
    float=htbp,
    caption={Parallel Interleaving Example},
    label={cs:interleaving_example},
    belowcaptionskip=0.5em,
    captionpos=b,
    language=Modest,
    style=ModestStyle
]
int x[] = {0, 0};
process setTo5(int id) {{= x[id] = 5; =}}
par { :: setTo5(0) :: setTo5(1) }
\end{lstlisting}

To resolve this issue we use the synchronizing actions provided by \modest. Actions provide a synchronization method where parallel assignments happen in lockstep with one other. For a more in-depth explanation of actions in \modest, see \cite{hartmanModestTutorial}. Adding a \lstinline{clock} action that models the synchronous nature of digital circuits eliminates states \(s_1\) and \(s_2\), and we are left with a single state transition that acts akin to the synchronous digital system as shown in Figure~\ref{subfig:action_ss}.
An alternative would be to apply probabilistic partial order reduction~\cite{BDG06,BGC04,DN04} to pick \emph{one interleaving} of the independent assignments; by exploiting the system's synchronous nature, we achieve a stronger reduction that collapses the entire interleaving into \emph{one transition} synchronizing all the involved parallel processes.

\begin{lstlisting}[
    float=htbp,
    caption={Synchronizing Parallel Interleavings with a Clock Action},
    aboveskip=2ex,
    label={cs:clock_action},
    captionpos=b,
    language=Modest,
    style=ModestStyle
]
action clock; int x[] = {0, 0};
process setTo5(int id) { clock{= x[id] = 5; =} }
par { :: setTo5(0) :: setTo5(1) }
\end{lstlisting}

\begin{figure}[hbt]
    \centering
    \includegraphics[%
        alt={Example trace of routers updating in sync}%
        ]{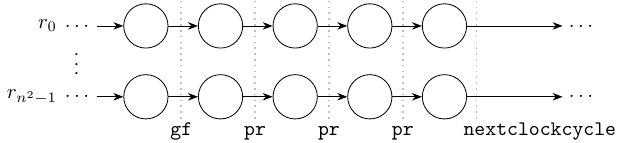}
    \caption{Impact of Synchronizing Actions in a Parallel Composition of \(n^2\) Routers}
    \label{fig:exampleofsyncstates}
\end{figure}

To synchronize the parallel composition of router processes, each assignment in each sub-process called in \lstinline{router} has a unique synchronizing action associated to it. This ensures that in each state update the synchronous semantics of digital circuits is correctly represented and, like Figure~\ref{fig:action_reduce_ss} shows, helps to reduce the state space and improve verification time. Code Segment~\ref{cs:proc:Router} shows the relevant sub-processes to be synchronized. Code Segment~\ref{cs:proc:allsync} shows an example of how each sub-process of \lstinline{Router} uses a unique action to ensure that the assignments are synchronized. The effect of these actions is that each of the routers update their state together, and no interleaving is allowed. Figure~\ref{fig:exampleofsyncstates} shows an example trace of the state updates of an \nxn{n} NoC, where each router updates in sync with the other routers.

\begin{lstlisting}[
    float=htbp,
    caption={Use of Synchronizing Actions in Router Processes},
    label={cs:proc:allsync},
    captionpos=b,
    language=Modest,
    style=ModestStyle,
    columns=fixed
]
action gf, pr, ar, up;
process GenerateFlits(int id) { gf {= /*...*/ =} }
process PrepRouter(int id)    { pr {= /*...*/ =} }
process AdvanceRouter(int id) { ar {= /*...*/ =} }
process UpdatePiority(int id) { up {= /*...*/ =} }
\end{lstlisting}

\subsection{Clock Process}
\label{sec:modular:clock}

A clock process (Code Segment \ref{cs:proc:Clock}) is composed with the router instances to count the elapsed clock cycles (Code Segment \ref{cs:Par}). The clock count is used to generate the PSN results (Section~\ref{sec_conclusions}) and to give context to different flit injection strategies. \modest allows for variable assignment to occur with a synchronization action (i.e. \lstinline{nextClockCycle}), which is used to increment the clock count.

\begin{lstlisting}[
    float=htbp,
    caption={\lstinline{Clock} Process in \modest},
    label={cs:proc:Clock},
    captionpos=b,
    language=Modest,
    style=ModestStyle
]
process Clock() { nextClockCycle{= clk = (clk + 1) =}; Clock()}
\end{lstlisting}

\subsection{Flit Generation Algorithm}
\label{sec:modular:flit_gen}

The modular \lstinline{GenerateFlits} process allows for easy customization of network traffic patterns in order to measure the PSN impact of different flit injection strategies. This process has one input parameter \lstinline{id} and may add a flit to \rib{id}{L}. This simulates the generation of network traffic in the NoC. Determining the traffic generation pattern needs to be done after observing patterns in real-world or simulated environments, as the generation pattern will vary based on application and use-case.

Results in this work use the periodic flit generation pattern from \cite{Roberts2021} with a 3/10 duty cycle. In this pattern, each router has a new flit synchronously added to the local buffer during the first 3 out of every 10 clock cycles. Destinations for each new flit are chosen according to a uniform distribution of all other routers in the NoC. While this specific distribution and generation pattern were chosen for direct comparison with \cite{Roberts2021}, the modular framework allows arbitrary generation patterns and destination distributions. Appendix~\ref{app:sec:processes:generate_flits} gives further examples.

\subsection{Routing Algorithm}
\label{sec:modular:flit_propogate}

The \textit{routing algorithm} used by each router determines how incoming flits should be treated. This is another modular process with input parameter \lstinline{id} that can be adjusted to model different routing algorithms in the NoC. Due to the page limit, the full details and semantics of the \lstinline{AdvanceRouter} process are included in the Appendix~\ref{app:sec:processes:advance_router}.

This work implements the X-Y routing algorithm, as described in Section~\ref{sec_preliminaries}, because it is deterministic and deadlock free. X-Y routing was used in \cite{Roberts2021,Lewis2019} and a goal of this work was to reproduce results from those papers. Additionally, 7 of the 16 real-world NoCs surveyed in~\cite{jerger_onchip_2017} used X-Y (or Y-X) routing.

\subsection{Priority Tracking and Updates}
\label{sec:modular:priority}

During routing, flits are passed between buffers on channels. Since a channel can be used \emph{only once} per clock cycle, conflicts may arise when multiple flits need to use the same channel. To ensure that all buffers are \textit{serviced}, flags are set when a buffer cannot be serviced due to conflicts. A buffer is \textit{serviced} when a flit is removed from the front and removed or sent down a channel to another buffer of the neighboring router.

The \lstinline{UpdatePriority} sub-process uses these flags to generate the priority of buffers to service for the next cycle. The order of buffers to be serviced next is stored in an array and used during the next clock cycle as the initial order for flits to be sent. As part of the modularity of this design, the priority algorithm used for each buffer can be reconfigured in the model to meet the design goals. This work uses a round-robin process to ensure that each buffer is serviced fairly.

\subsection{Noise Tracking}
\label{sec:modular:noise}

PSN is characterized by tracking the activity level of each router in the network every clock cycle. As discussed in Section~\ref{sec:prelim:psn}, resistive noise is related to the activity during the current clock cycle, and inductive noise is related to the change in activity over two consecutive clock cycles. Further discussion of PSN characterization is given in Section~\ref{sec_psn_prob}.

\section{Correctness of the Modular NoC}
\label{sec_verification}

\modest uses the NoC model we provide to explore the complete state-space, and then the \modestToolset's \lstinline{mcsta} model checker~\cite{HartmannsH15} uses \ensuremath{\forall\Box} and $\exists\Diamond$ CTL formulas to verify the correctness of a NoC design. Verification of these properties is carried out on a generic router and inter-router communications. In this section, $i$ and $j$ represent valid router IDs in a NoC and \(\gen{j}{i}\) denotes a packet destined for \(j\) being added to the \(Local\) buffer of \(\ri{i}\).

\subsection{Flit Generation Verification}

As discussed in Section \ref{sec:modular:flit_gen}, the flit generation pattern for a given NoC design can be easily modified. Thus verifying a unique flit generation pattern requires a verification strategy suited to each pattern, so we do not present a general method for verifying the correctness of \emph{all} flit generation patterns.

However, one property that all flit generation patterns must demonstrate across all paths is that a flit destined for the router they are originating from should not be generated as shown in Property~\ref{eq:prop:noFlitsForSelf}.

\begin{equation}
    \forall i:\forall\Box(\neg (\gen{i}{i}))
    \label{eq:prop:noFlitsForSelf}
\end{equation}

While we do not present a method for verifying all possible flit generation patterns, we do demonstrate a possible approach to such verification. The flit generation algorithm used in \cite{Roberts2021} assigns destinations using a uniformly random distribution. We can verify that there exists an execution path in which a flit is sent from router $i$ to router $j$ for all routers $i,j$ using the following CTL property:

\begin{equation}
    \forall i:\forall j : j \neq i \implies \exists\Diamond(\gen{j}{i})
    \label{eq:prop:equalDestinations}
\end{equation}

\subsection{Scheduling Priority Verification}

The implementation of this scheduler relies on a list of buffers that have been serviced or are waiting for service. It is important that this list is coherent, i.e.\ that the list always contains all buffers and no duplicate entries. Property~\ref{eq:prop:priorityList} shows the CTL formula for this property, where $i$ is a placeholder for router ID.

\begin{equation}
\begin{aligned}
    \forall i: \forall\Box&(
        \mathit{local}\in\ri{i}.\mathit{priorityList}         \land \mathit{north}\in\ri{i}.\mathit{priorityList} \\
        &\land\: \mathit{east}\in\ri{i}.\mathit{priorityList} \land \mathit{south}\in\ri{i}.\mathit{priorityList} \\
        &\land \mathit{west}\in\ri{i}.\mathit{priorityList})  \land \mathrm{length}(priorityList) = 5
\end{aligned}
\label{eq:prop:priorityList}
\end{equation}

\subsection{Flit Propagation Verification}

The real-world NoC design deploys buffers of bounded sizes, but they are modeled as an unbounded queue in \modest. This design choice is used to facilitate the instantiation of a buffer of arbitrary size, but makes it possible for a bounded buffer to exceed its specified size in the model. Therefore, we must specify a property to check that all the buffers in a NoC never exceed the user-specified size. This property should also check that a flit is never accidentally sent to a full buffer, as that would increase the length of the buffer by one.
In CTL, this is
\begin{equation}
    \forall i, j : \forall\Box(\mathrm{length(}\ri{i}.\mathit{buffer}_{j}) \leqslant \mathit{BUFFER\_SIZE})
    \label{eq:prop:bufferLength}
\end{equation}

Additionally, each channel should only admit at most one flit during a clock cycle to correctly model a synchronous NoC. We verify this property by adding a counter to each channel that tracks how many times it was used during a single clock cycle. It is incremented when a flit is sent across the channel during the \lstinline{AdvanceRouter} process, and is reset back to zero in the \lstinline{UpdatePriority} process. The CTL property to check that no counter is ever greater than one is
\begin{equation}
    \forall i, j : \forall\Box(\ri{i}.\mathit{channel}_{j}.\mathit{used\_count} \leqslant 1)
    \label{eq:prop:channelUsedOnce}
\end{equation}

\subsection{Improvement Over Previous Works}

Previous works such as \cite{Roberts2021,boe2023probabilistic} were not able to perform CTL model checking as \lstinline{mcsta} did not support CTL then. Instead, they encoded functional correctness using the PCTL syntax and checked if the probability of PCTL encodings of Properties~\ref{eq:prop:noFlitsForSelf}-\ref{eq:prop:channelUsedOnce} were $0.0$ or $1.0$, which may not be the same for certain probabilistic loops, and requires less scalable model checking methods.

Additionally, due to the many unnecessary interleavings generated by the lack of true synchronization as discussed in Section~\ref{sec:modular:router}, previous models were only able to be checked up to a small number of clock cycles, effectively performing bounded model checking. Our work checks an unbounded number of cycles for the \nxn{2} NoC. Checking \nxn{3} however still runs out of memory; future work can look at abstracting parts of the modular model to scale up verification.

\section{PSN Probability}
\label{sec_psn_prob}

As discussed in Section~\ref{sec:prelim:psn}, this work measures PSN through an behavioral description instead of circuit level details such as resistance, inductance, and current. This section outlines our method for calculating PSN probabilities and how it is customized to provide more information.

Resistive noise occurs when activity occurs in a router during a single clock cycle. Each router maintains a local \lstinline{activity} counter to track events that occur during a single clock cycle. The counter \lstinline{activity} increases when a packet is sent or received. This activity level translates into a resistive noise event when it crosses the activity threshold $K$ (i.e., \lstinline{ACTIVITY_THRESH}) set by the user. We then use Property \ref{eq:res} to determine the cumulative probability that $N$ resistive noise events occur in $k$ clock cycles. $N$ is set by the user as \lstinline{RESISTIVE_THRESH}.

Inductive noise occurs when the activity level in a router changes between two clock cycles. Each router also maintains a local \lstinline{lastActivity} variable that stores the activity of the previous clock cycle. When the difference between \lstinline{activity} and \lstinline{lastActivity} is greater than the activity threshold $K$, it generates an inductive noise event. We then use Property \ref{eq:ind} to determine the cumulative probability that $M$ inductive noise events occur. $M$ is set by the user as \lstinline{INDUCTIVE_THRESH}.

By default, the PSN characterization in this model is a global characterization, where PSN is considered for all routers in the network. However, since each router maintains a local \lstinline{activity} and \lstinline{lastActivity} variable, it is easy to extract PSN probabilities for a single router or group of routers instead of all routers, as shown in Section~\ref{sec_results}.

\section{Results and Discussion}
\label{sec_results}

The results presented in this work were generated on a machine using a 12-core AMD Ryzen Threadripper Processor (at 3.5 GHz), 132 GB of memory, and Ubuntu Linux version 22.04 LTS. \modestToolset version 3.1.290 was used. All formal models, plots, and scripts are publicly available at~\cite{noc_github}. The runtime results are presented in Table \ref{tbl:RuntimeResults}.

\begin{table}[htb]
    \centering
    \begin{tabular}{ccccccc}
        \hline
        \textbf{Size} & \textbf{Parameter} & \textbf{Tool} & \textbf{Confidence} & \textbf{Runtime} & \textbf{Memory} & \textbf{States} \\
        & & & \textbf{Interval} & (HH:MM:SS) & (MB) & \\
        \hline
        \nxn{2} & Functional Properties  & \texttt{mcsta} & --     & 00:10:31 & 9844 & 3446073 \\
        \nxn{2} & Resistive PSN          & \texttt{modes} & $95\%$ & 00:06:11 & 105 & -- \\
        \nxn{2} & Inductive PSN          & \texttt{modes} & $95\%$ & 00:56:08 & 105 & -- \\
        \nxn{3} & Resistive PSN          & \texttt{modes} & $95\%$ & 00:03:07 & 129 & -- \\
        \nxn{3} & Inductive PSN          & \texttt{modes} & $95\%$ & 00:14:13 & 131 & -- \\
        \nxn{4} & Resistive PSN          & \texttt{modes} & $95\%$ & 00:02:42 & 164 & -- \\
        \nxn{4} & Inductive PSN          & \texttt{modes} & $95\%$ & 00:16:13 & 172 & -- \\
        \nxn{8} & Resistive PSN          & \texttt{modes} & $95\%$ & 00:05:23 & 545 & -- \\
        \nxn{8} & Inductive PSN          & \texttt{modes} & $95\%$ & 00:21:08 & 565 & -- \\
        \hline
        \textbf{Total} &                 &                &        & 02:14:36 & & \\
        \hline
    \end{tabular}
    \vspace{1mm}
    \caption{Runtime Results}
    \label{tbl:RuntimeResults}
\end{table}

\subsection{%
    \texorpdfstring%
    {CTL Verification Results for a \nxn{2} NoC}%
    {CTL Verification Results for a 2x2 NoC}%
}
\label{sec:results:ctl_2x2}

CTL model checking results for the \nxn{2} NoC were generated using \lstinline{mcsta}, the model checking engine in \modest. In addition, the \lstinline{chainopt} option was used to reduce the state space by collapsing chains of states connected by probability-1 transitions. All properties specified in Section \ref{sec_verification} were verified on the \nxn{2} model.

\subsection{PSN Results}
\label{sec:results:psn}

The results of the PSN PCTL properties were captured using \lstinline{modes}, the statistical model checker in \modest. SMC was used due to memory constraints for models larger than \nxn{2}. In \lstinline{modes}, the \lstinline{max-run-length} option is set to zero to allow longer simulation runs (as \lstinline{modes} by default aborts when it encounters a simulation run longer than 10,000 transitions, as a heuristic to avoid hanging on unintended infinite paths -- which our model does not have).

The main goal of this work was to create a modular NoC framework that is more flexible and scalable than the previous work. However, in order to reproduce and compare the probabilistic verification results between this work and similar previous work~\cite{Roberts2021}, we configure the NoC model to match the previous ones. Therefore, this model uses a buffer length of four, the X-Y routing algorithm, an activity threshold of three, and the 3/10 flit injection pattern (detailed in Section~\ref{sec:modular:flit_gen}) unless otherwise stated.

\subsubsection{%
    \texorpdfstring%
    {\nxn{2} NoC Verification Results}%
    {2x2 NoC Verification Results}%
}
\label{sec:2x2_results}

\begin{figure}[hbt]
    \centering%
    \subfloat[][{Inductive Noise}]{%
        \centering%
        \includegraphics[%
            alt={2x2 resisitive PSN CDF},%
            width=0.5\linewidth%
            ]{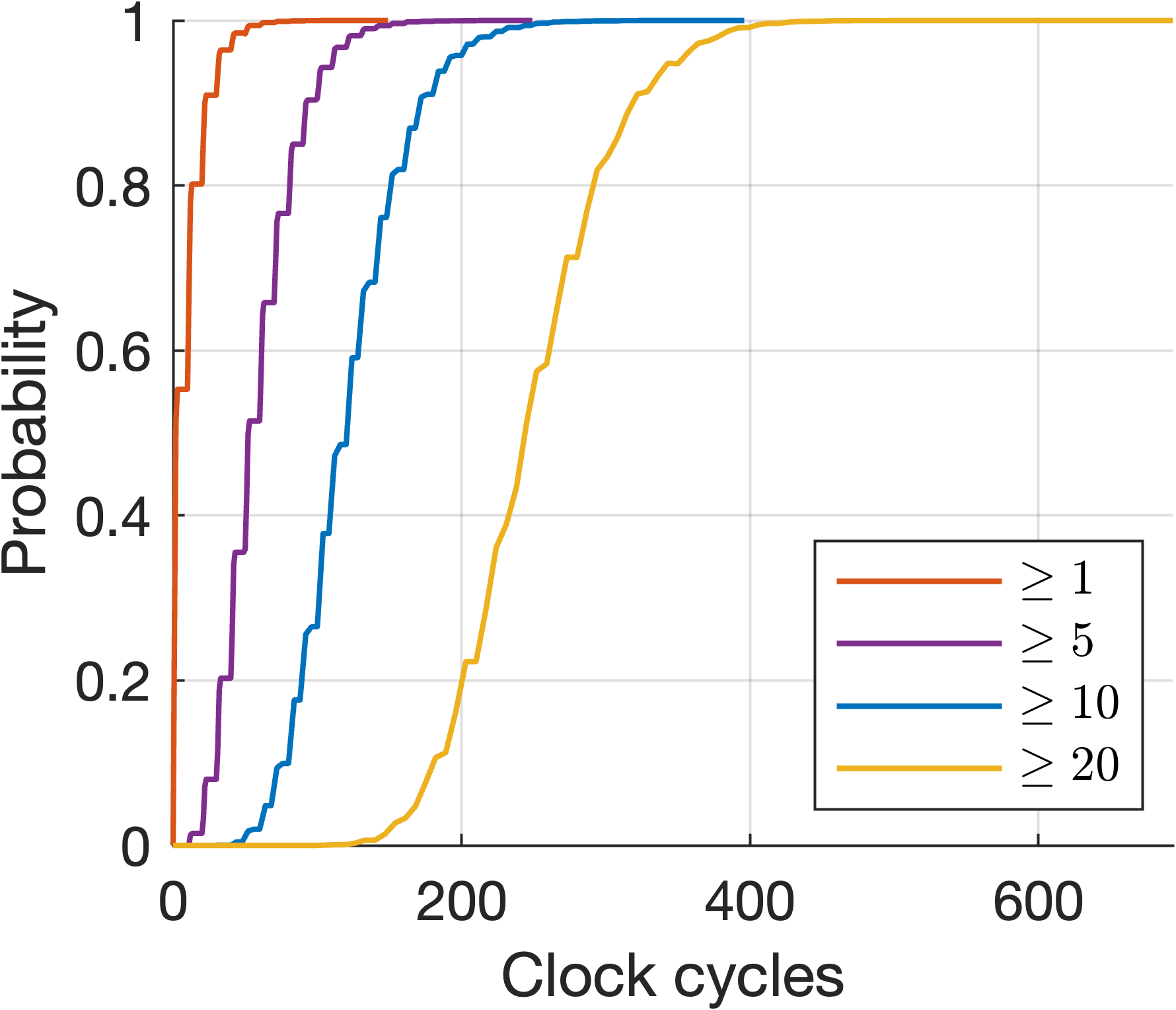}%
        \label{subfig:2x2ResistFinal}%
    }%
    \subfloat[][{Inductive Noise}]{%
        \centering%
        \includegraphics[%
            alt={2x2 inductive PSN CDF},%
            width=0.5\linewidth%
            ]{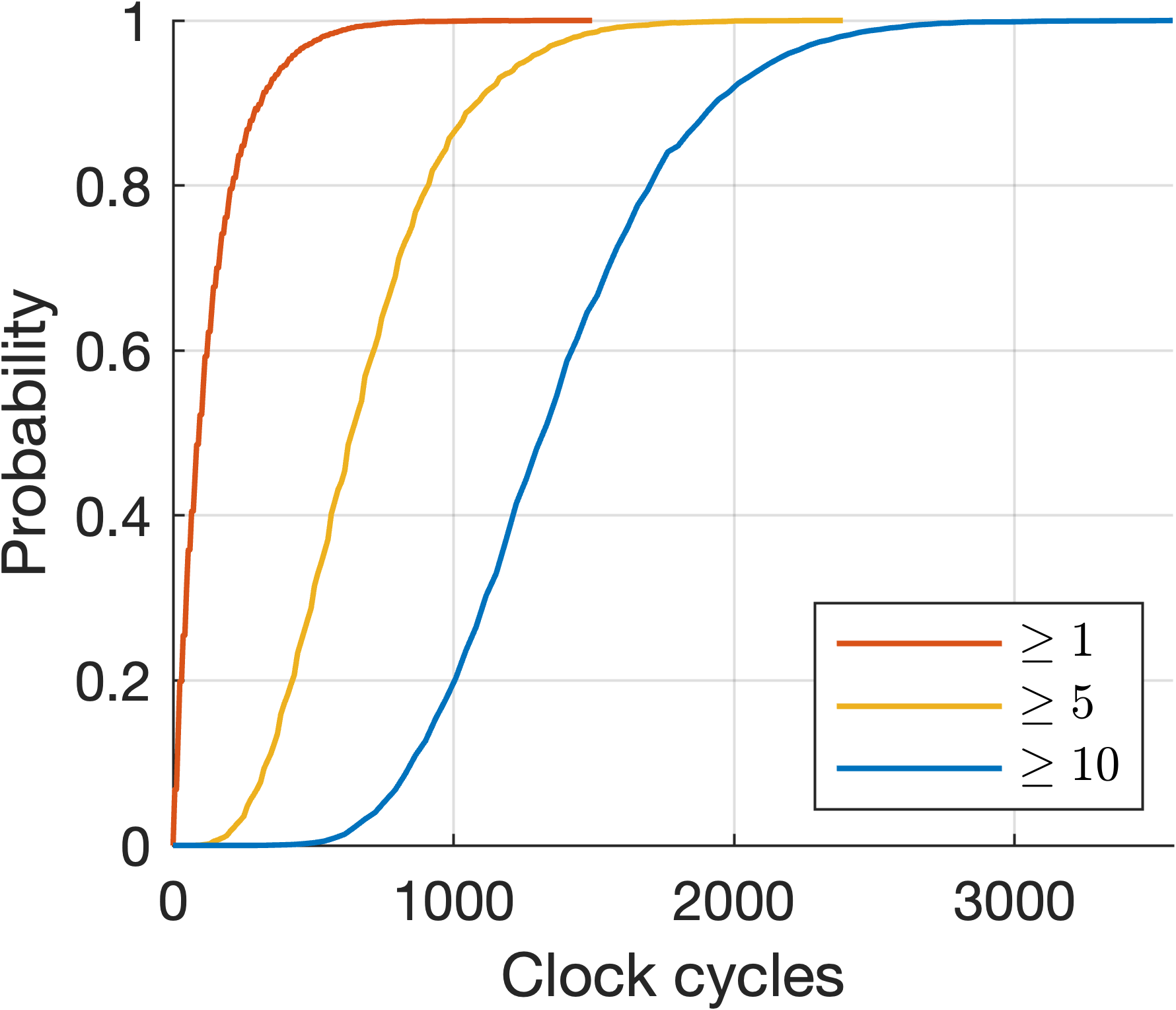}%
        \label{subfig:2x2InductFinal}%
    }%
    \caption{PSN for \nxn{2} Modular Model}
    \label{fig:2x2FinalResults}
\end{figure}

Figure~\ref{fig:2x2FinalResults} shows the PSN results for a \nxn{2} NoC for our modular design. Each plot shows the \emph{cumulative distribution functions} (CDFs) of the probability of PSN events greater than or equal to the amount shown in the legend. Both resistive and inductive noise display a step-like growth corresponding to the 3/10 flit injection pattern detailed in Section~\ref{sec:modular:flit_gen}.

\subsection{%
    \texorpdfstring%
    {Comparison to Previous \nxn{2} NoC Model}%
    {Comparison to Previous 2x2 NoC Model}%
}
\label{sec:comparison}

The \nxn{2} model results in~\cite{Roberts2021} were regenerated using the \lstinline{modes} SMC tool and compared the modular \nxn{2} results of this work in order to reproduce their findings. A similar comparison of a central router from~\cite{Lewis2019} to the modular model is given in~\cite{boe2023probabilistic}.

The initial results of the modular design were quite different from \cite{Roberts2021}. Because the modular model had been proved functionally correct by passing the CTL properties detailed in Section~\ref{sec_verification}, we determined the discrepancies derived either from architectural differences or from an error in previous work.

We used comparative model checking to narrow down the differences between the two models and discovered two architectural differences. The initial modular implementation did not generate new flits on clock cycle 0, while the previous monolithic model of \cite{Roberts2021} did. We corrected this by rearranging the sub-processes shown in Code Segment \ref{cs:proc:Router} to occur before the synchronizing clock action.

Additionally, the modular model allows multiple flits to be consumed during a single clock cycle, while the monolithic model allowed only a single flit to be consumed per cycle. For example, if two flits destined for $\ri{0}$ arrived at $\ri{0}$ on the same clock cycle in different channels, the modular model would consume both flits, while the monolithic model would consume only one. While both designs are valid, the reality is that the hardware attached to each router is unknown. It is therefore \emph{assumed} in this work that the hardware attached to each router is capable of consuming multiple flits per cycle, and we adjusted the monolithic model to accept multiple flits per clock cycle.

\begin{figure}[hbt]
    \centering%
    \subfloat[][{Before Comparative Model Checking}]{%
        \centering%
        \includegraphics[%
            alt={Differing output before comparative model checking},%
            width=0.5\linewidth%
            ]{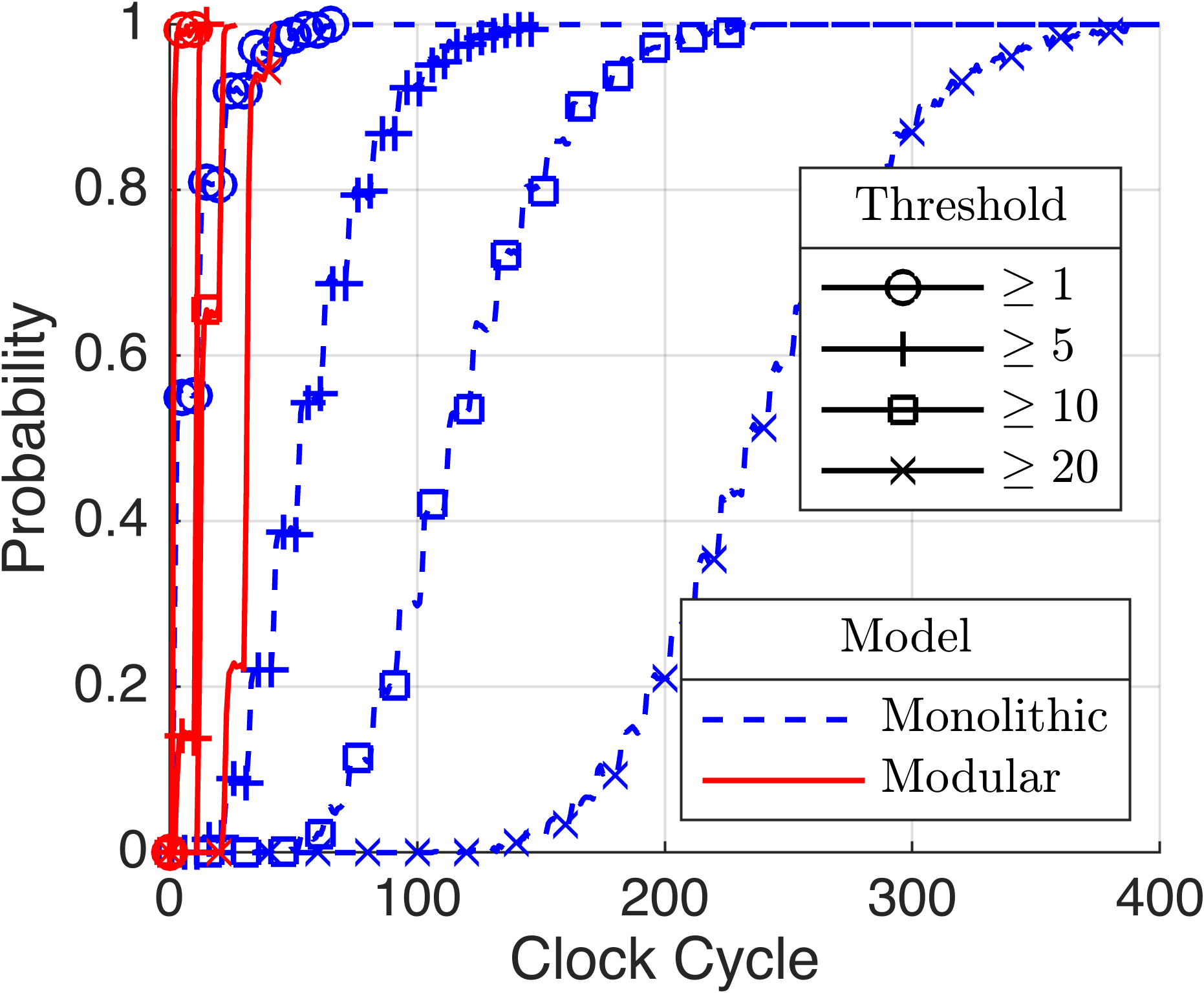}%
        \label{fig:ResistComp}%
    }%
    \subfloat[][{After Comparative Model Checking}]{%
        \centering%
        \includegraphics[%
            alt={Matching output after comparative model checking},%
            width=0.5\linewidth%
            ]{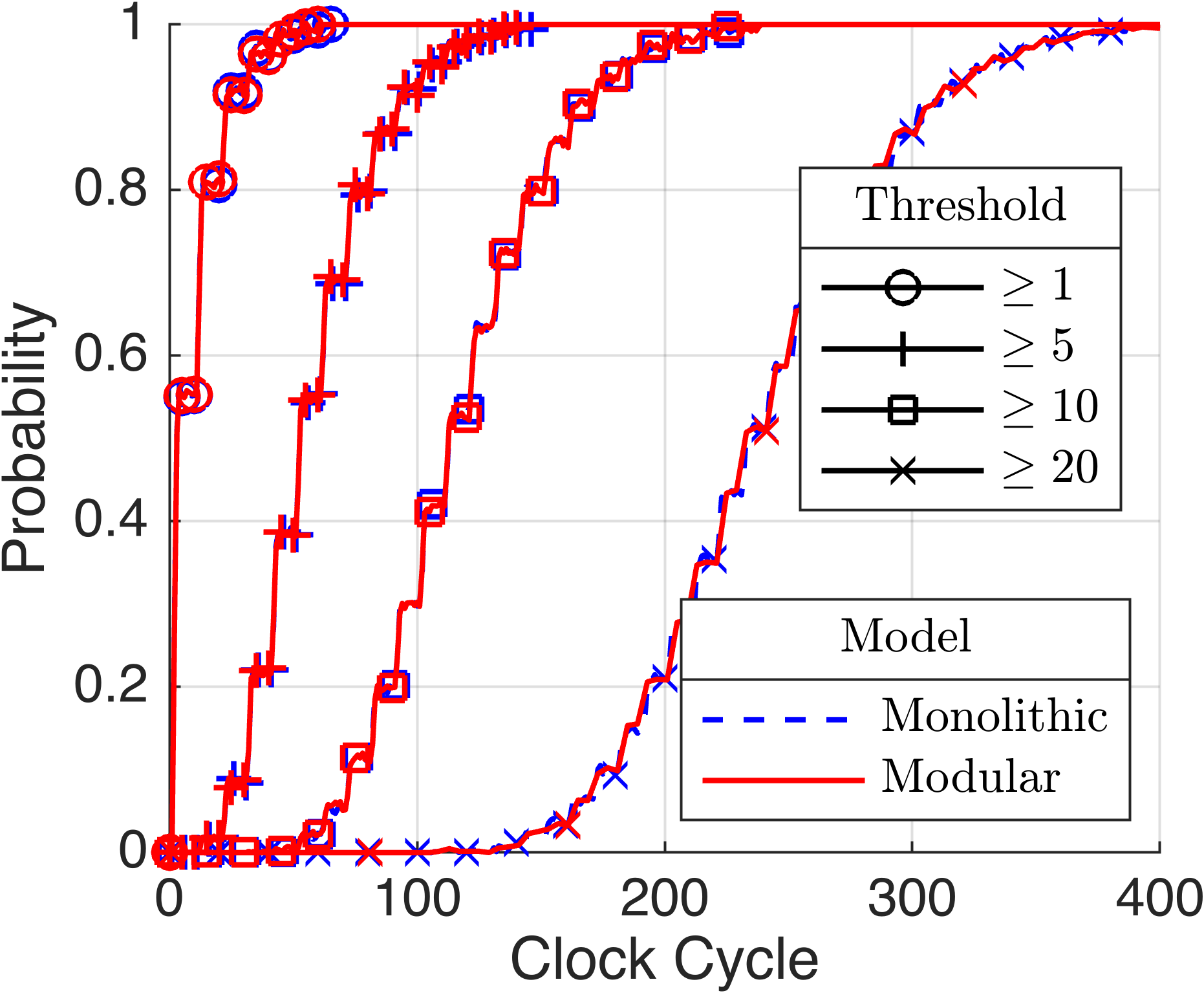}%
        \label{fig:InductComp}%
    }%
    \caption{Modular and Monolithic \nxn{2} NoC Resistive Noise Comparison}
    \label{fig:TotalComp}
\end{figure}

We performed comparative model checking by assuming that both models were correct, and then generating the PSN characterization for each model and comparing the output. If the output differs, at least one of the models is faulty as they should be equivalent. Determining which model is faulty is done by simulating traces where the correct output is known on each model. Whichever model returns the incorrect output is the faulty model, and the trace can be analyzed to determine where the error originated. This process is repeated iteratively until the models match. We used this method because we could not directly compare states due to the different implementations of each model. Matching results after comparative-model checking was completed is shown in Figure \ref{fig:TotalComp}.

\subsection{PSN Results for Larger NoCs}

\begin{figure}[hbt]
    \centering%
    \subfloat[][{\nxn{3}}]{%
        \centering%
        \includegraphics[%
            alt={3x3 inductive PSN CDF},%
            width=0.33\linewidth%
            ]{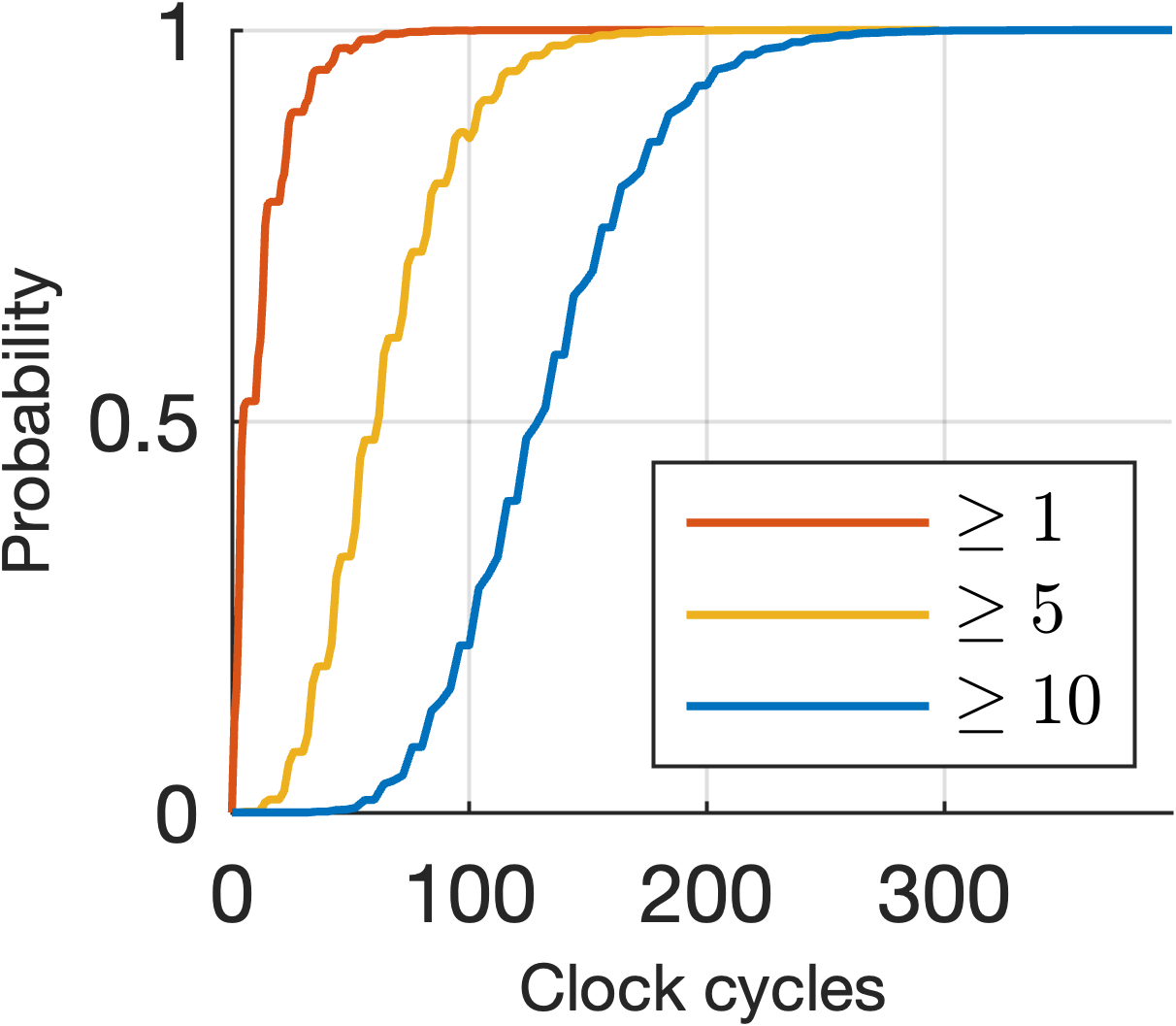}%
        \label{fig:3x3Induct}%
    }%
    \subfloat[][{\nxn{4}}]{%
        \centering%
        \includegraphics[%
            alt={4x4 inductive PSN CDF},%
            width=0.33\linewidth%
            ]{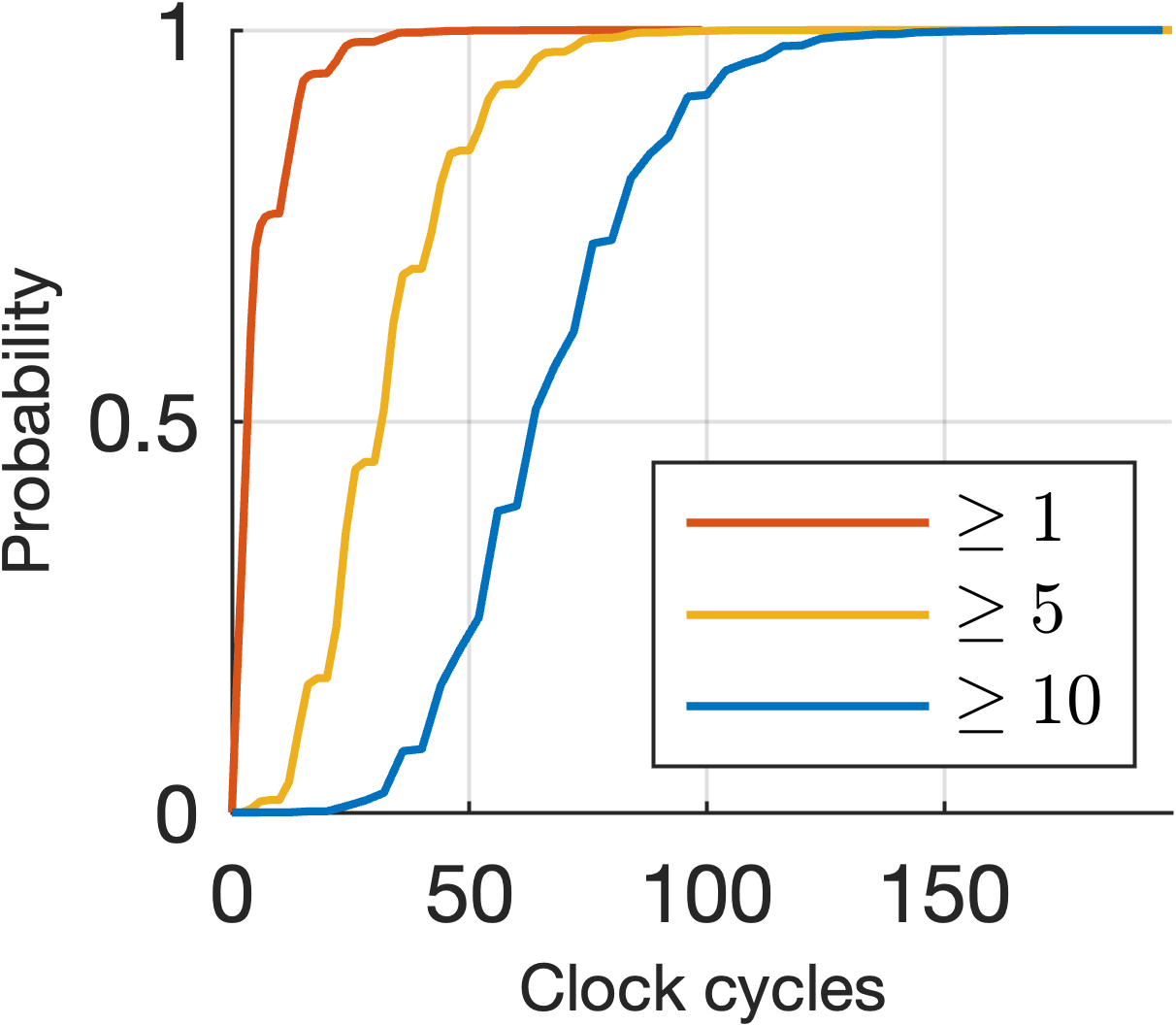}%
        \label{fig:4x4Induct}%
    }%
    \subfloat[][{\nxn{8}}]{%
        \centering%
        \includegraphics[%
            alt={8x8 inductive PSN CDF},%
            width=0.34\linewidth%
            ]{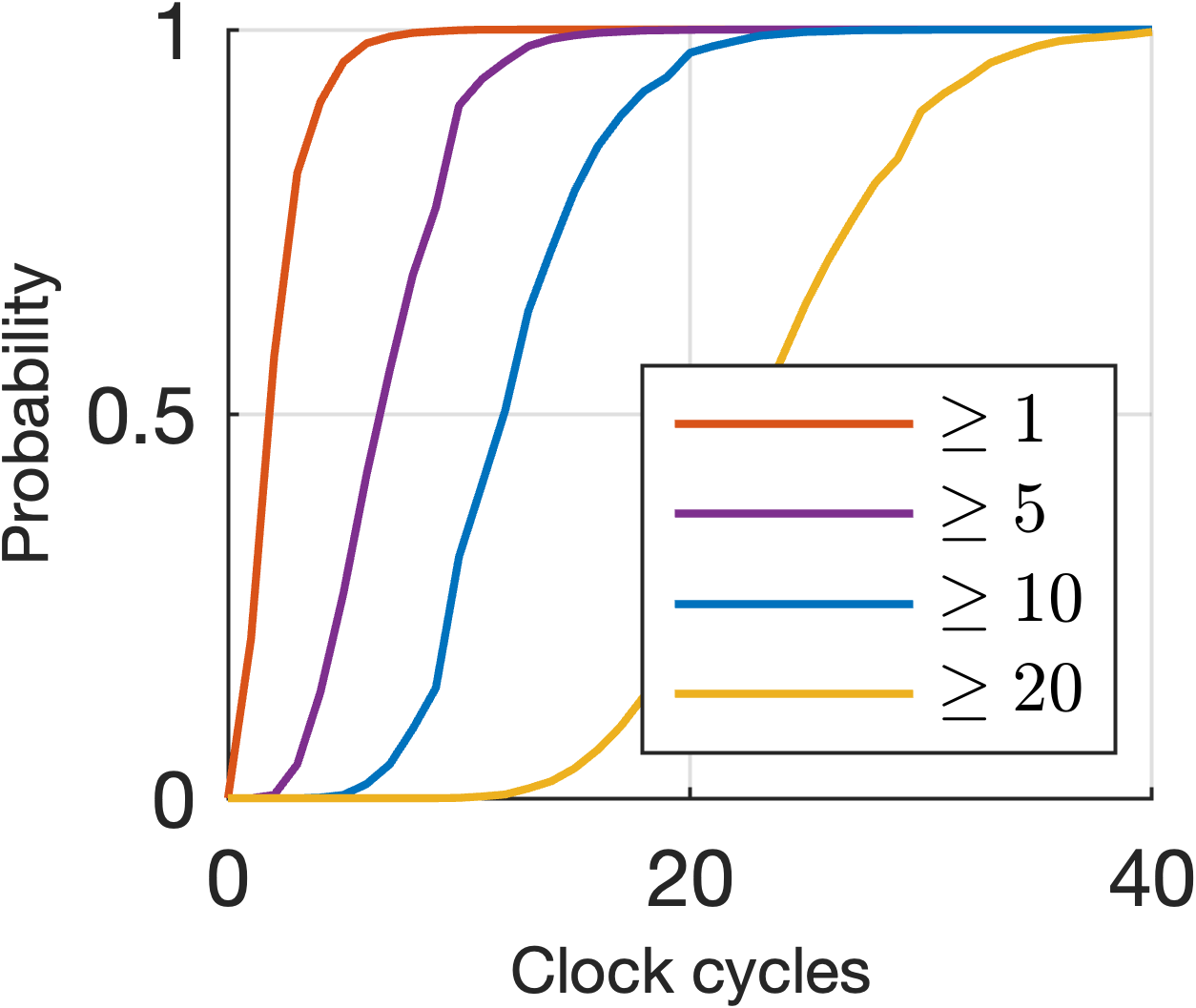}%
        \label{fig:8x8Induct}%
    }%
    \caption{Inductive Noise CDF for \nxn{3}, \nxn{4}, and \nxn{8} Modular Model}
    \label{fig:3x3_4x4_8x8Results}
\end{figure}

The advantage of a modular NoC model is the ability to scale it with ease. By instantiating 9, 16, and 64 routers for \nxn{3}, \nxn{4}, and \nxn{8} NoCs, respectively, we are quickly able to scale our PSN analysis. Figure \ref{fig:3x3_4x4_8x8Results} shows the inductive noise CDFs for these sizes. Resistive noise CDFs are not shown here due to page limitations, but are shown in Appendix~\ref{app:sec:additional_results}.

Similar to the \nxn{2} results in Figure \ref{fig:2x2FinalResults}, the \nxn{3} and \nxn{4} results show a pronounced pattern of steep slopes followed by rough edges which is a result of our chosen 3/10 flit injection pattern. As expected, compared to a \nxn{2} NoC, the likelihood of PSN events is much higher and occurs sooner in the \nxn{3} and \nxn{4} NoCs due to the higher number of routers and network traffic. Inductive noise events are likely to happen sooner in the \nxn{8} NoC compared to smaller models, and the pronounced steps every 3/10 cycles are gone. This is because packets are much more likely to persist in the \nxn{8} NoC, as packets in the \nxn{8} are likely to travel farther.

\begin{figure}[hbt]
    \centering%
    \subfloat[][{Resistive Noise}]{%
        \centering%
        \includegraphics[%
            alt={3x3 resistive PSN CDF per router},%
            width=0.5\linewidth%
            ]{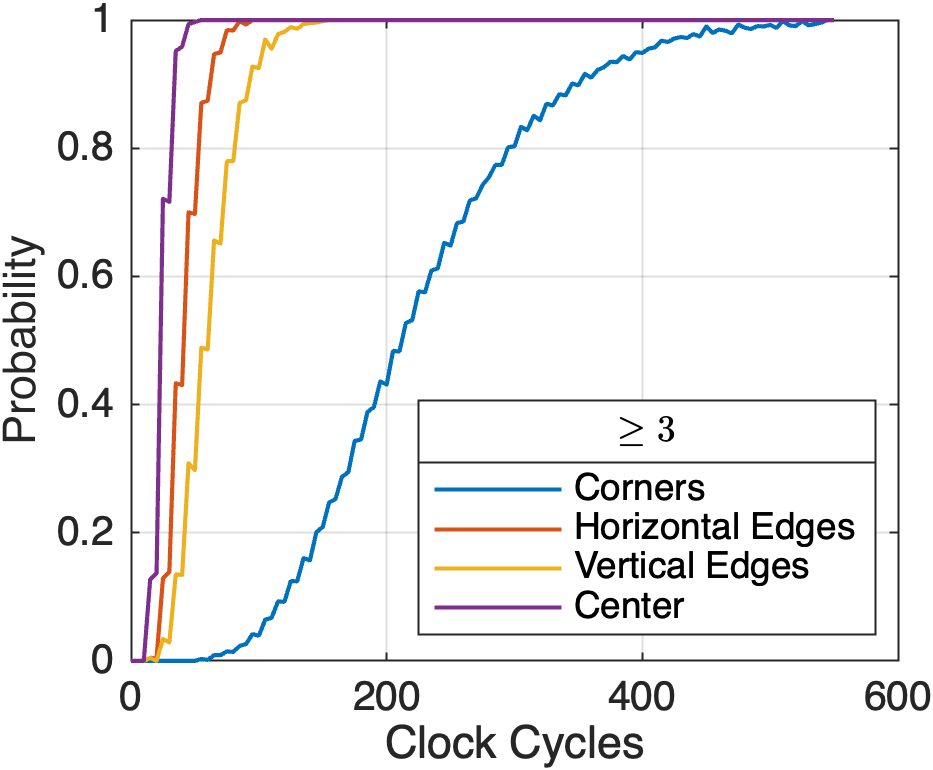}%
        \label{fig:3x3ResistDifferentNodes}%
    }%
    \subfloat[][{Inductive Noise}]{%
        \centering%
        \includegraphics[%
            alt={3x3 inductive PSN CDF per router},%
            width=0.5\linewidth%
            ]{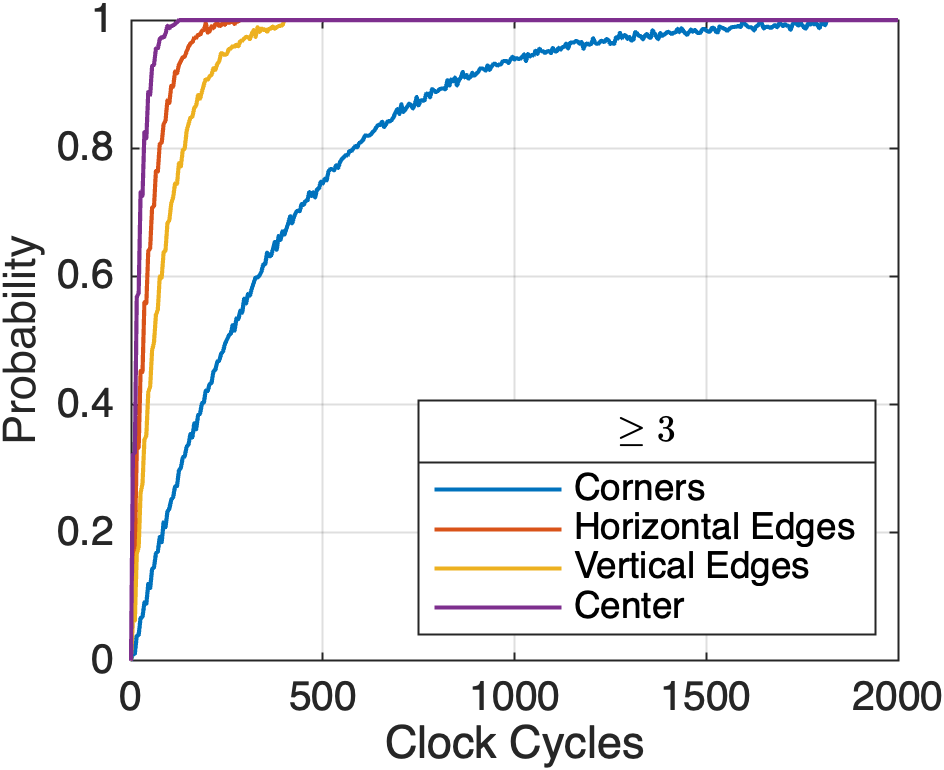}%
        \label{fig:3x3InductDifferentNodes}%
    }%
    \caption{Router Specific PSN for \nxn{3} Modular Model}
    \label{fig:3x3RouterSpecificNoise}
\end{figure}

\subsection{PSN for Specific Routers}

One difference between a \nxn{2} and \nxn{3} NoC are the different routing connections used. A \nxn{2} NoC, such as the one shown in Figure~\ref{fig:2x2config}, only uses \textit{corner routers} with two neighbors. A \nxn{3} configuration has \textit{corner routers} with two neighbors, \textit{edge routers} with three neighbors, and a \textit{central router} with four neighbors. 

The modular NoC model allows us to obtain insights that were impossible to obtain from the monolithic design style \cite{Roberts2021,Lewis2019}. Rather than checking PSN globally, we can inspect the \lstinline{activity} and \lstinline{lastActivity} of an individual router to get a CDF for PSN for that specific router. The results of PSN resulted from different router types are shown in Figure \ref{fig:3x3RouterSpecificNoise}. PSN results were generated for each router type with an \lstinline{ACTIVITY_THRESH} of 3.

The router-specific results in Figure~\ref{fig:3x3RouterSpecificNoise} show that the PSN increases toward the center of the network. Although the corner routers experienced a very gradual increase in PSN with respect to time, the center router experienced high levels of PSN almost immediately. This is because the central router has more active buffers than the others, and consequently, it experiences higher PSN. However, the horizontal edge routers experienced a higher PSN than those on the vertical edges, regardless of having the same number of buffers. This is a result of the X-Y routing causing more traffic through those areas. Therefore, placing most traffic-prone neighbors on the NoC perimeter may reduce PSN as the effects of PSN are less prevalent on corner and edge routers.

This ability to check PSN for specific parts of a NoC can be used to provide valuable information on how the chosen routing algorithm and the flit injection pattern affect PSN. For example, if a new routing algorithm was designed to minimize traffic to the central router of a \nxn{3} NoC, our modular model could verify the effects on PSN that the new algorithm would have compared to the current X-Y routing algorithm. This would allow routing algorithm design to occur early in the design cycle, leading to more effective designs.

Previously, the largest topology achieved for a probabilistic NoC model was a \nxn{2} network. As mentioned, \nxn{2} routers would only have three input buffers (two neighbor buffers and a local buffer). With the PSN threshold set to three, this inherently results in lower PSN between routers with higher conflict rates, because if even one buffer goes unserviced in a cycle, there can be no PSN on that router. PSN characterization for \nxn{2} networks does not scale to larger NoCs, making it difficult to make any particular recommendation to chip designers based on \nxn{2} findings alone.

In the work of~\cite{Roberts2021}, it is concluded that PSN can be reduced by injecting delays between flits. While this is true, this work affirms that PSN is difficult to completely eradicate using the flit injection pattern alone. The larger a NoC becomes, the longer it takes on average for flits to propagate through the system. Injecting flits in only a single clock cycle can still result in PSN. In addition to reducing the flit injection rate, this work recommends grouping routers into high-traffic regions, with the highest traffic connections reserved for vertical neighbors. In this way, flits can propagate through the NoC more quickly. This recommendation, in conjunction with keeping heavy traffic on the perimeter of the NoC, may result in even lower traffic through the central routers.

\section{Conclusion}
\label{sec_conclusions}

This paper presents a case study on the development of an easy-to-use, scalable, and modular formal NoC model using the \modest modeling language. The paper describes the CTL properties to verify the correctness of the model and the PCTL properties to quantify PSN. It also describes the diagnoses of a discrepancy in the quantification of PSN in a \nxn{2} NoC while attempting to reproduce a previous work~\cite{Roberts2021}, which revealed an inaccurate modeling of the flit consumption behavior in~\cite{Roberts2021}. In addition, results from statistical model checking on the \nxn{3}, \nxn{4}, and \nxn{8} NoCs suggested network flit scheduling schemes to minimize PSN. While this work focuses on NoCs sized up to \nxn{8}, it can be easily scaled to larger models. The modular design of the model makes it ideal for examining NoC designs early in the design cycle. Doing so may help find potential errors sooner and may encourage designs that are more robust in their function. Future work includes scaling the CTL correctness verification to arbitrarily sized models and exploring a wide range of flit injection patterns and routing algorithms.
\section*{Data Availability Statement}
\label{sec:dataavailability}

A docker environment with the tools, models, and results of this paper is available on \href{https://doi.org/10.5281/zenodo.17247418}{\texttt{Zenodo}}~\cite{waddoups_2025_17247418}. Additionally, the models are available on \href{https://github.com/formal-verification-research/VMCAI26_Modular_NoC_Artifact}{\texttt{GitHub}}~\cite{noc_github}.

\clearpage
\bibliographystyle{splncs04}
\bibliography{base}

\clearpage
\appendix
\chapter*{Appendix}

\section{Modular NoC Implementation}
\label{app:sec}

This section gives an outline of the NoC design implemented in \modest. For information on how to generate unique models, see Appendix~\ref{app:sec:python}, and for details on the each process in the model see Appendix~\ref{app:sec:processes}. Additional results are shown in Appendix~\ref{app:sec:additional_results}.

\subsection{Additional Results}
\label{app:sec:additional_results}

\begin{figure}[hbt]
    \centering%
    \subfloat[][{Resistive Noise}]{%
        \centering%
        \includegraphics[width=0.5\linewidth]{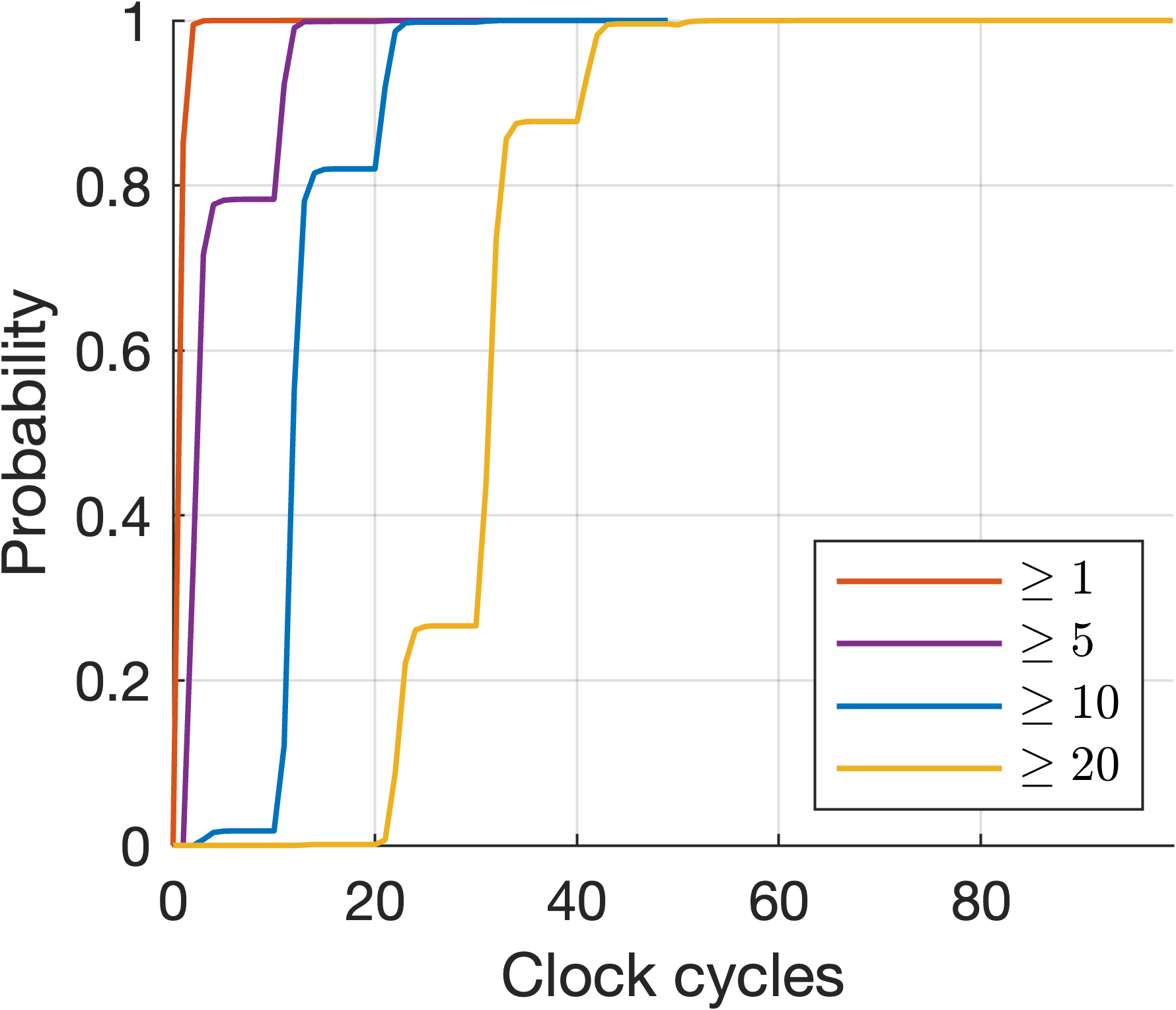}%
    }%
    \subfloat[][{Inductive Noise}]{%
        \centering%
        \includegraphics[width=0.5\linewidth]{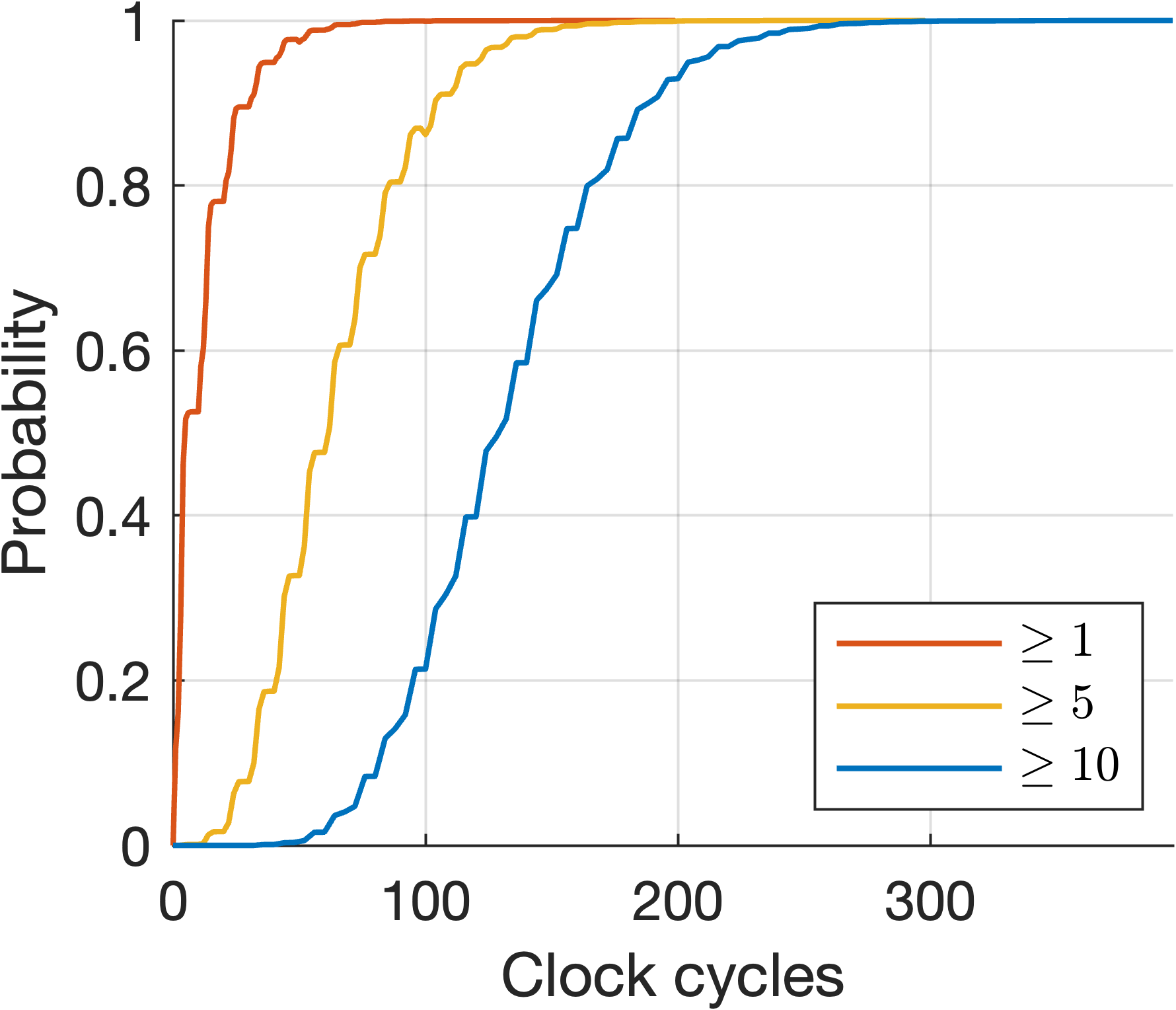}%
    }%
    \caption{PSN CDF for \nxn{3} Modular Model}
    \label{app:fig:3x3results}
\end{figure}

\begin{figure}[hbt]
    \centering%
    \subfloat[][{Resistive Noise}]{%
        \centering%
        \includegraphics[width=0.5\linewidth]{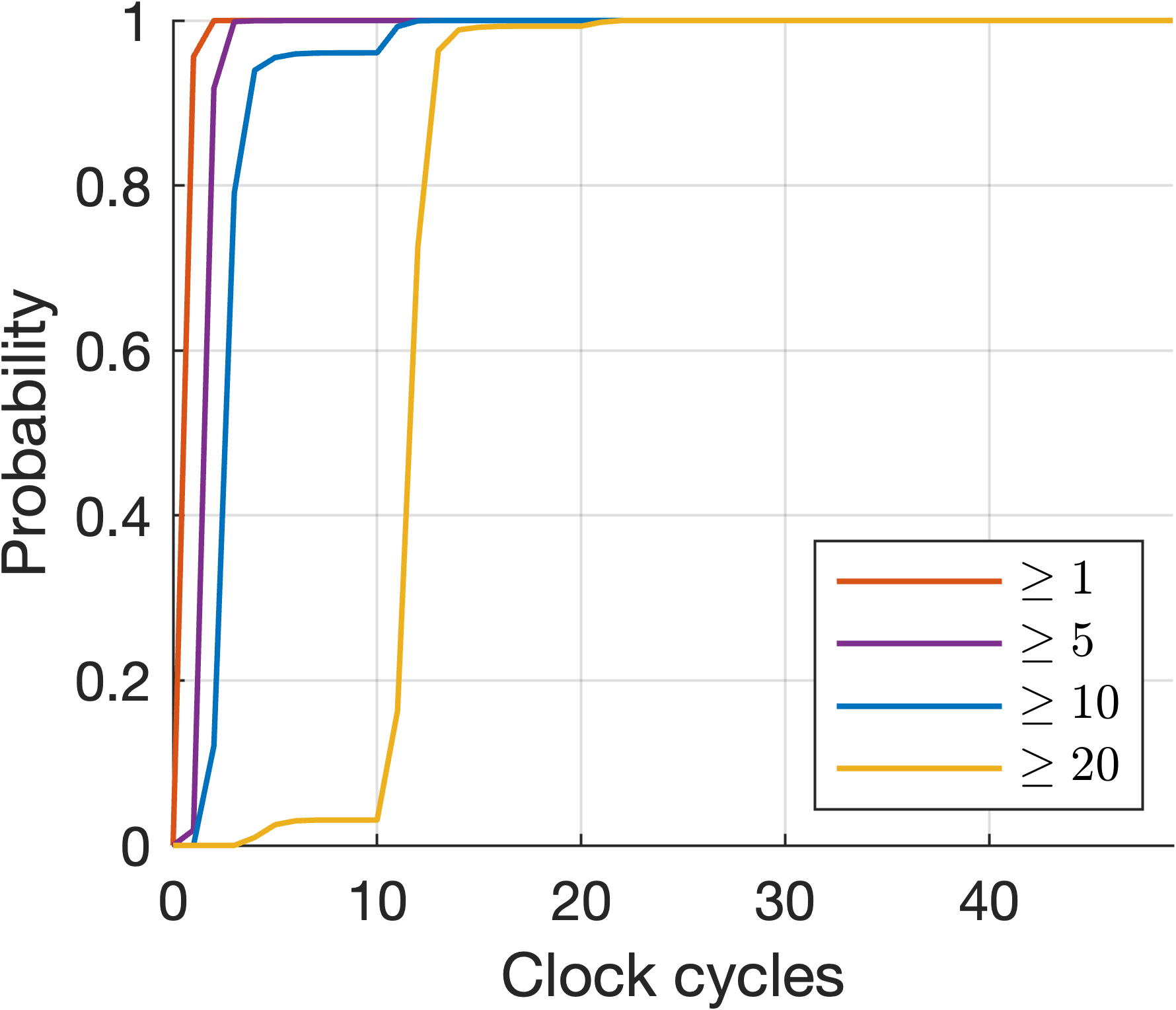}%
    }%
    \subfloat[][{Inductive Noise}]{%
        \centering%
        \includegraphics[width=0.5\linewidth]{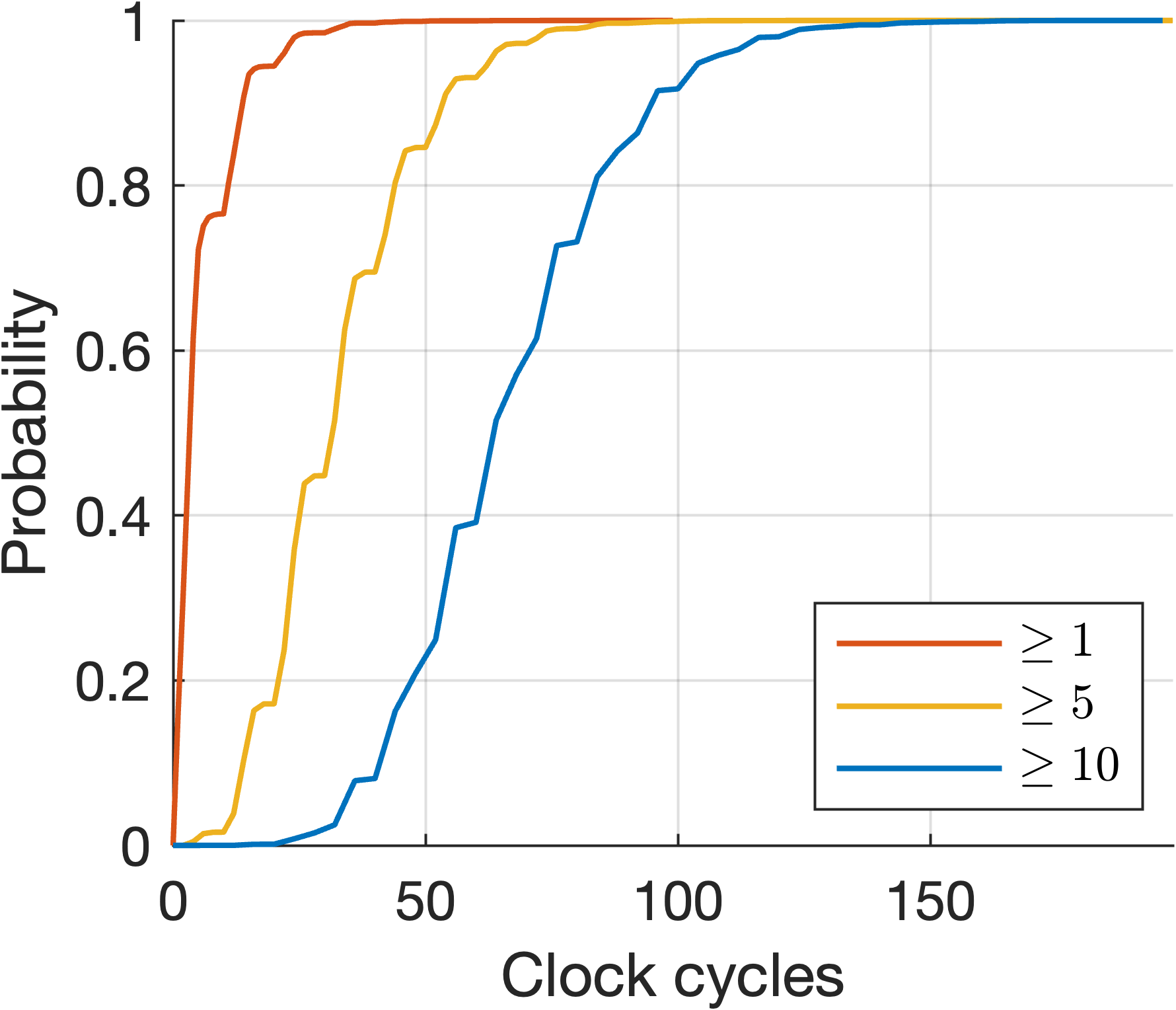}%
    }%
    \caption{PSN CDF for \nxn{4} Modular Model}
    \label{app:fig:4x4results}
\end{figure}

\begin{figure}[hbt]
    \centering%
    \subfloat[][{Resistive Noise}]{%
        \centering%
        \includegraphics[width=0.5\linewidth]{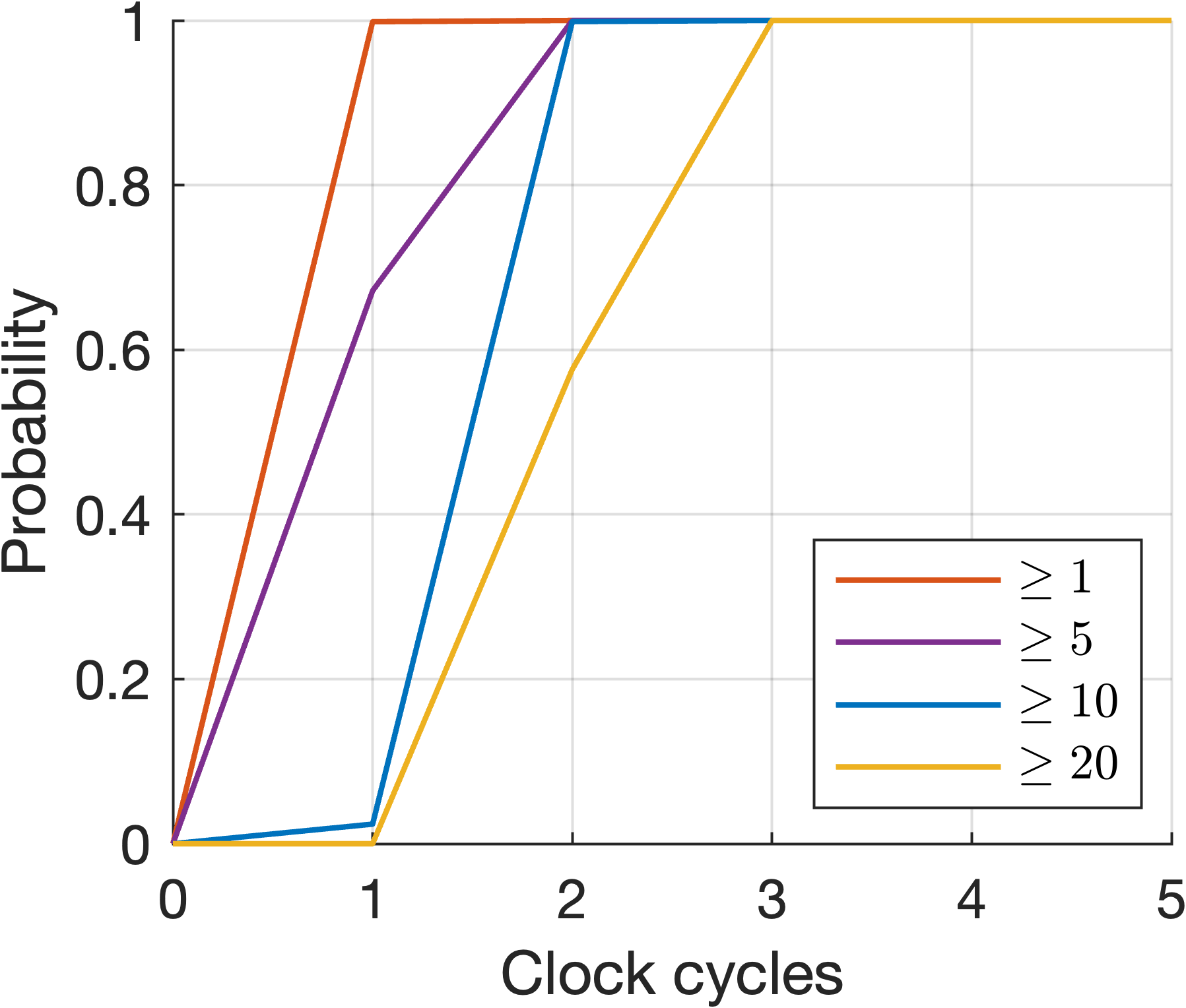}%
    }%
    \subfloat[][{Inductive Noise}]{%
        \centering%
        \includegraphics[width=0.5\linewidth]{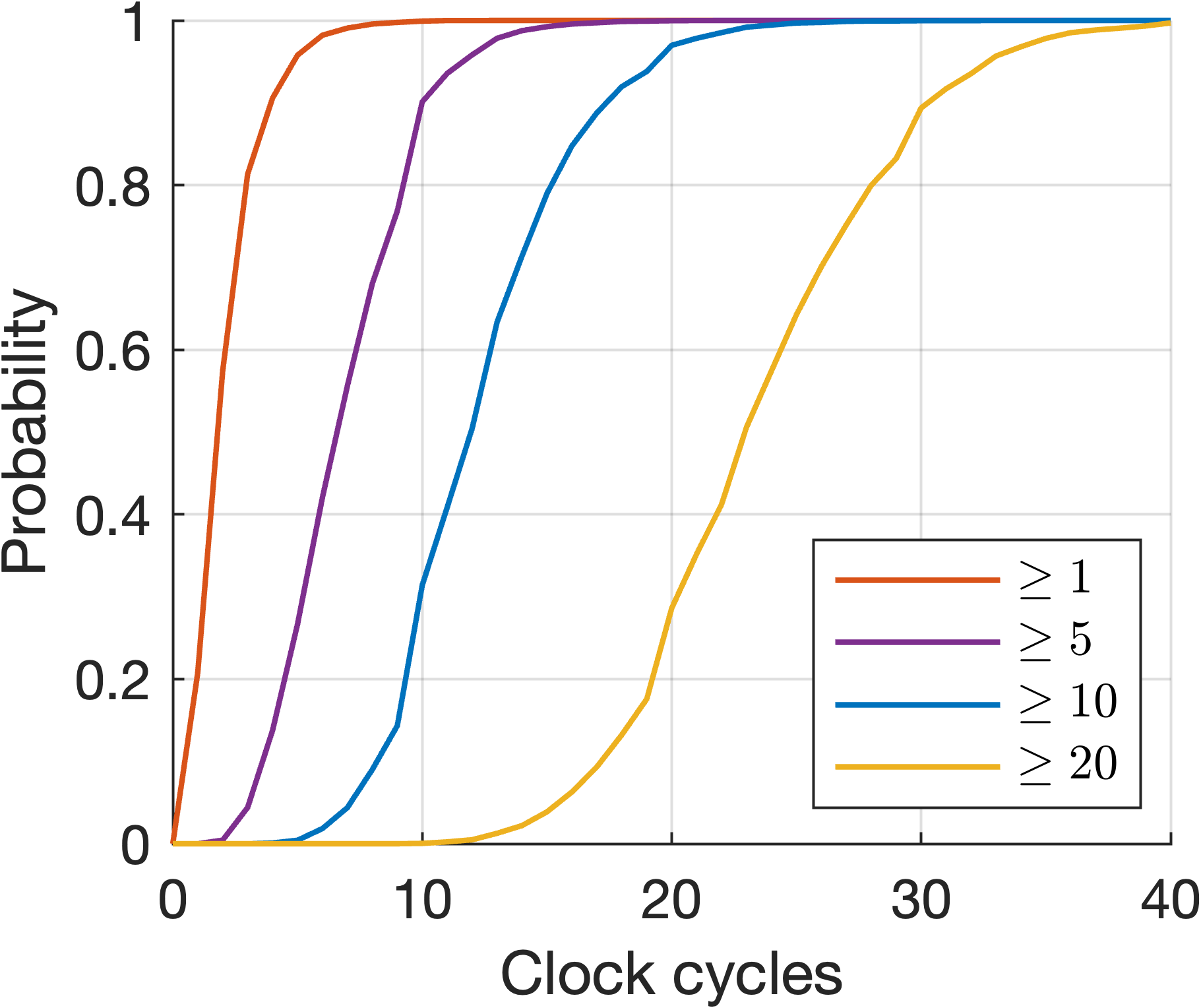}%
    }%
    \caption{PSN CDF for \nxn{8} Modular Model}
    \label{app:fig:8x8results}
\end{figure}

Additional results are shown in Figures~\ref{app:fig:3x3results},~\ref{app:fig:4x4results}, and~\ref{app:fig:8x8results}. These results are all generated using the statistical model checker \lstinline{modes} in \modest, and use the same configuration as the results in Section~\ref{sec_results} (buffer size of four, 3/10 flit generation pattern, round robin scheduling, XY routing).

Similar to the analysis done in Section~\ref{sec_results}, we can see that the probability of reaching $K$ PSN events grows with NoC size, as expected. Additionally, for the \nxn{3} and \nxn{4} the step pattern corresponding to the 3/10 flit generation pattern is clearly visible. For \nxn{8} results the step pattern is not pronounced, as it's more likely that the destination for a flit will be further away, thus keeping flits in the network for longer. Additionally, in an \nxn{8} NoC we observe that within three clock cycles we are likely to see 20 or more resistive noise events. This speaks to the large number of central routers in an \nxn{8} NoC that each have four neighbors, and as such are likely to have higher average traffic than corner or edge routers.

\subsection{Datatypes}
\label{app:sec:types}

\modest allows for the creation of high-level datatypes to represent state. We create several custom datatypes to define the structure and state of the NoC.

\subsubsection{Router Datatype}
\label{app:sec:types:router}

The router datatype defined in Code Segment~\ref{app:cs:router_type} holds the state for a single router. This includes each buffer and channel, as well as data related to the arbitration algorithm (Section \ref{sec:modular}), and variables for tracking the number of PSN events (\lstinline{activity}). In \modest, the syntax  \lstinline{int(A..B)} defines an integer type in $[A,B]$.

\begin{table}[h!]
    \centering
    \caption{Description of router member variables}
    \label{tab:router_variables}
    \begin{tabular}{|l|l|}
        \hline
        \textbf{Member Variable}    & \textbf{Description} \\ \hline
        \texttt{channels}           & The buffer and channel instances. \\ \hline
        \texttt{ids}                & The IDs of neighboring routers. \\ \hline
        \texttt{priority\_list}     & A priority queue of serviced routers \\ \hline
        \texttt{priority\_list\_temp} & A temp queue used during calculation \\ \hline
        \texttt{serviced\_index}    & \\ \hline
        \texttt{unserviced\_index}  & \\ \hline
        \texttt{total\_unserviced}  & \\ \hline
        \texttt{thisActivity}       & Activity counter for current clock cycle \\ \hline
        \texttt{lastActivity}       & Activity counter for last clock cycle \\ \hline
        \texttt{used}               & \\ \hline
    \end{tabular}
\end{table}

\begin{lstlisting}[
    float=htbp,
    caption={\lstinline{router} Datatype},
    label={app:cs:router_type},
    belowcaptionskip=0.5em,
    captionpos=b,
    language=Modest,
    style=ModestStyle
]
datatype router = {
    channel[] channels,
    int(-1..NOC_MAX_ID)[] ids,
    int(0..4)[] priority_list,
    int(0..4)[] priority_list_temp,
    int(0..4) serviced_index,
    int(0..4) unserviced_index,
    int(0..5) total_unserviced,
    int thisActivity,
    int lastActivity,
    bool[] used
};
\end{lstlisting}

\subsubsection{Channel Datatype}
\label{app:sec:types:channel}

The channel datatype defined in Code Segment~\ref{app:cs:channel_type} holds the information needed for correctly modeling a buffer in \modest. The \lstinline{option} modifier in \modest declares a type that either holds a value or is empty.

\begin{lstlisting}[
    float=htbp,
    caption={\lstinline{channel} Datatype},
    label={app:cs:channel_type},
    belowcaptionskip=0.5em,
    captionpos=b,
    language=Modest,
    style=ModestStyle
]
datatype channel = {
    buffer option buffer,
    bool serviced,
    bool isEmpty,
    bool isFull
};
\end{lstlisting}

\subsubsection{Buffer Datatype}
\label{app:sec:types:buffer}

The buffer datatype shown in Code Segment~\ref{app:cs:buffer_type} is modeled as  linked list, where there is some data stored with an optional tail pointing to the next element in the list. In this work, the data we store in the buffer is the same as the data in a packet in the network. Our packets only carry the ID of the destination router. The destination ID is valid from $[0,n^2-1]$ (Section \ref{sec:modular}) and we model that using a bounded int type in \modest.

\begin{lstlisting}[
    float=htbp,
    caption={\lstinline{buffer} Datatype},
    label={app:cs:buffer_type},
    belowcaptionskip=0.5em,
    captionpos=b,
    language=Modest,
    style=ModestStyle
]
datatype buffer = {
    int(0..NOC_MAX_ID) hd,
    buffer option tl
};
\end{lstlisting}

\subsection{NoC State}
\label{app:sec:noc_state}

The NoC state is stored as an array of router instances, where the index of each router instance corresponds to the ID of router. This state is stored as a global variable because each process needs to be able to access this state. The syntax for the \lstinline{noc} router is shown in Code Segment~\ref{app:cs:noc_structure}.

\begin{lstlisting}[
    float=htbp,
    caption={NoC Data Structure},
    label={app:cs:noc_structure},
    belowcaptionskip=0.5em,
    captionpos=b,
    language=Modest,
    style=ModestStyle
]
router[] noc = [ /*... */ ]
\end{lstlisting}

Global arrays in \modest do not need to initialized at the start of model, however some of the CTL properties we specify in Section \ref{sec_verification} must hold for all states, including the initial state. If the \lstinline{noc} array is not default initialized, then the CTL properties fail on the initial state. Because of this, we explicitly initialize the \lstinline{noc} array in the initial state as shown in Code Segment~\ref{app:cs:noc_structure_init}.

\begin{lstlisting}[
    float=htbp,
    caption={Default Initialization},
    label={app:cs:noc_structure_init},
    belowcaptionskip=0.5em,
    captionpos=b,
    language=Modest,
    style=ModestStyle
]
router[] noc = [
router {
    channels: [
        channel {isEmpty: true, isFull: false},
        channel {isEmpty: true, isFull: false},
        channel {isEmpty: true, isFull: false},
        channel {isEmpty: true, isFull: false},
        channel {isEmpty: true, isFull: false}],
    ids: [NO_CONNECT, NO_CONNECT, 1, 2],
    priority_list: [NORTH, EAST, SOUTH, WEST, LOCAL],
    priority_list_temp: [0, 0, 0, 0, 0],
    serviced_index: 0,
    unserviced_index: 0,
    total_unserviced: 0,
    thisActivity: 0,
    lastActivity: 0,
    used: [false, false, false, false, false]
},
/* ... continued for each router ... */ ]
\end{lstlisting}
Other variables that track the state of the NoC are \lstinline{resistiveNoise} and \lstinline{resistiveNoise} counters. These counters track the total number of PSN events that happen in the system. Additionally, there is a integer \lstinline{clk} that tracks the number of elapsed clock cycles in the system. This counter is necessary for measuring cycle-wise PSN, but must be removed during CTL model checking as we aim to check the system for unbounded clock cycles, as there are finite states but infinite clock cycles.

\subsection{Processes}
\label{app:sec:processes}

As described in Section \ref{sec:modular}, the NoC model is composed of modular processes that are composed together. Each process is modularized by taking an ID as a parameter. In each modular process, this ID is then used to read and update the router state in the \lstinline{noc} array. For example, in a \nxn{2} NoC four router processes are composed in parallel, with each given a unique ID as shown in Code Segment \ref{cs:Par}. In each router process this unique ID is used to read and update the state of the router located at \lstinline{noc[id]}.

\subsubsection{Router Process}
\label{app:sec:processes:router}

The router process (Code Segment~\ref{app:cs:router_process}) is the \textit{top-level} process in our hierarchical design. This process is made up of several sub-processes, each of which execute a step in the router model, composed together in serial. Functionality over each clock cycle is done by a recursive call to the router process after calling the \lstinline{nextClockCycle} action. Some sub-processes are described in further detail below, while others are commented in the code available in \cite{noc_github}.

\begin{lstlisting}[
    float=htbp,
    caption={\lstinline{Router} Process},
    label={app:cs:router_process},
    belowcaptionskip=0.5em,
    captionpos=b,
    language=Modest,
    style=ModestStyle
]
process Router(int id) {
    GenerateFlits(id);
    PrepRouter(id);
    AdvanceRouter(id);
    UpdatePiority(id);
    UpdateGlobalNoiseTracking(id);
    nextClockCycle;
    Router(id)
}
\end{lstlisting}

\subsubsection{Generate Flits Process}
\label{app:sec:processes:generate_flits}

As described in Section \ref{sec:modular:flit_gen}, the flit generation process can be adjusted to model unique flit generation algorithms. This process takes an ID, and can optionally add a new element to the local buffer of \ri{ID}, provided that the local buffer of \ri{ID} is not full.

Code Segment \ref{app:cs:generate_flit_process} shows the implementation of the 3/10 flit generation pattern, where a new flit is generated the first three out of every ten clock cycles. The basic outline of the process follows the template given in Code Segment~\ref{app:cs:generate_flits_template}, with one branch adding a new flit and the other branch doing nothing. \lstinline{id} is used to access a specific router instance in the \lstinline{noc} array. In each branch the \lstinline{generateFlits} synchronizing action is used to ensure that all state updates made by the \lstinline{GenerateFlits} processes are atomic across all routers. This synchronizing action is required to minimize the state space and prevent write-before-read conflicts, as described in Section \ref{sec:modular:sync_actions}.

\begin{lstlisting}[
    float=htbp,
    caption={\lstinline{GenerateFlits} Process Template},
    label={app:cs:generate_flits_template},
    belowcaptionskip=0.5em,
    captionpos=b,
    language=Modest,
    style=ModestStyle
]
process GenerateFlits(int id) {
    if (/* local buffer is not full */) { 
        /* conditionally add a new flit to the local buffer of r_id */
    } else {
        tau /* do nothing */
    }
}
\end{lstlisting}

In the 3/10 implementation shown in Code Segment \ref{app:cs:generate_flit_process}, we ensure Property \ref{eq:prop:noFlitsForSelf} from Section \ref{sec_verification} by providing a uniformly random destination id that is shifted at our current id so as to not generate a flit for the originating router, and to maintain a uniformly random destination address.

\begin{lstlisting}[
    float=htbp,
    caption={\lstinline{GenerateFlits} Process},
    label={app:cs:generate_flit_process},
    belowcaptionskip=0.5em,
    captionpos=b,
    language=Modest,
    style=ModestStyle
]
action generateFlits;
process GenerateFlits(int id) {
 int(0..NOC_MAX_ID) destination;

 if (!isBufferFull(noc[id].channels[LOCAL].buffer)
     && clk % 10 < 3) {
  generateFlits {=
   0: destination = DiscreteUniform(0, NOC_MAX_ID - 1),
   1: noc[id].channels[LOCAL].buffer = 
   enqueue(destination >= id ? 
           destination + 1 : 
           destination, noc[id].channels[LOCAL].buffer)
 =}}
 else { generateFlits }
}
\end{lstlisting}

Another possible implementation for the flit generation process is shown in Code Segment \ref{app:cs:generate_flit_bursty}. This implemenation is specifically designed for a \nxn{2} NoC and introduces two new pieces of global state: the \lstinline{burst_times} and \lstinline{sleep_times} arrays. These arrays are indexed by the ID of a router, and contain a clock cycle count. Both counts start at zero. On the first clock cycle, both \lstinline{burst_times[id]} and \lstinline{sleep_times[id]} are initialized to a discrete random value in the bounds set by \lstinline{[BURST_MIN:BURST_MAX]} and \lstinline{[SLEEP_MIN:SLEEP_MAX]}. These values are how many flits will be injected and how many clock cycles no new flits will be injected, respectively. This example uses the same uniform distribution of destination as the example in Code Segment \ref{app:cs:generate_flit_process}.

This process can simulate a router in a NoC where there are "bursty" periods of activity where 10-100 flits need to be sent, followed by "slow periods" where nothing is sent. As an example, if in clock cycle 1, \lstinline{burst_times[0]} is set to 20, and \lstinline{sleep_times[0]} is set to 250, then over the course of the next 270 cycles \ri{0} will have 20 new flits injected, followed by a period where no new flits are injected for 250 cycles -- assuming that the local buffer never gets full, in which case this whole period would be longer than 250 cycles.

\begin{figure}
    \centering
    \includegraphics[width=0.50\linewidth]{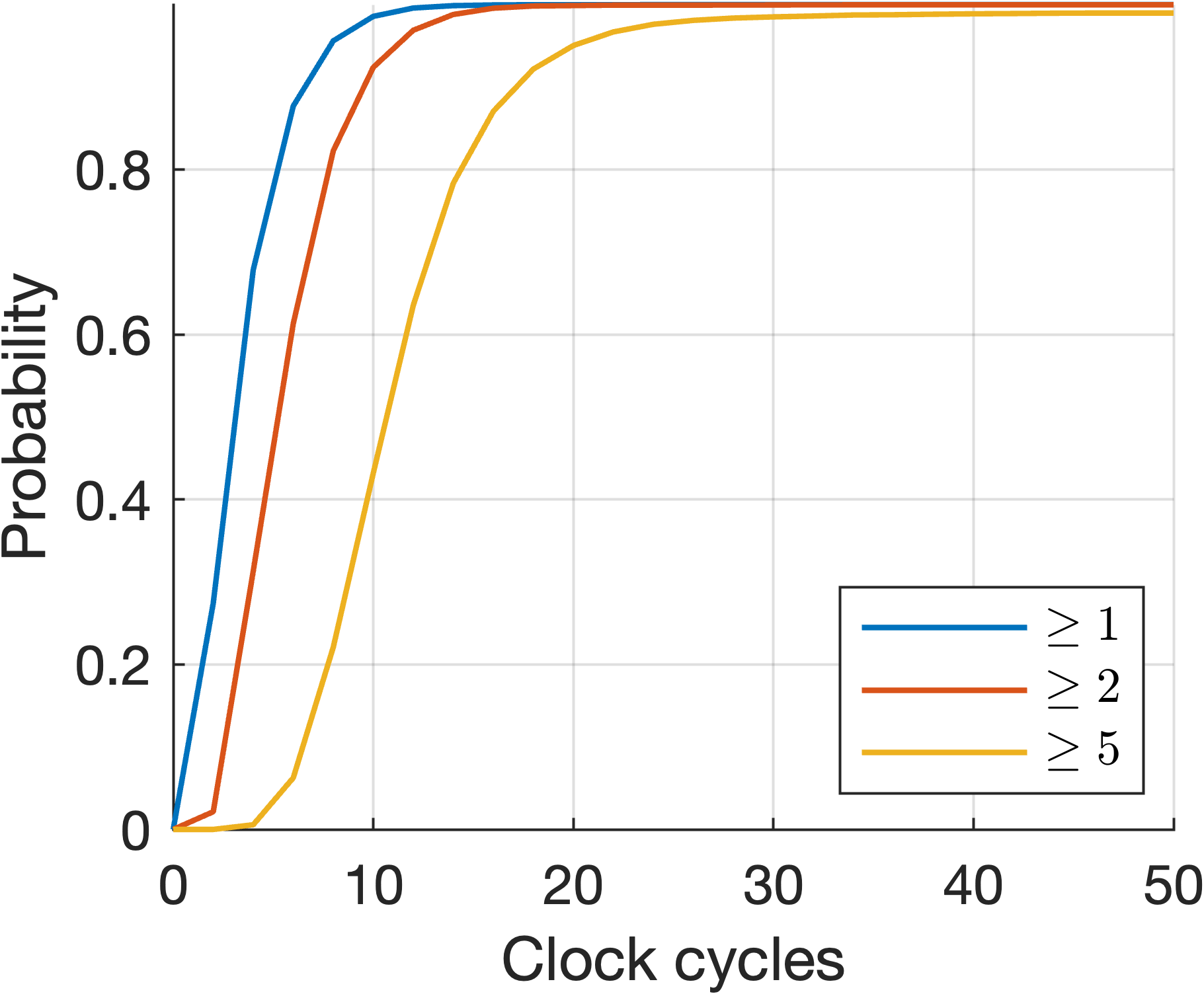}
    \caption{Resistive PSN CDF for \nxn{2} Modular Model with Code Segment~\ref{app:cs:generate_flit_bursty} Flit Injection Pattern}
    \label{app:fig:2x2_bursty_results}
\end{figure}

Results from this "bursty" flit injection pattern are shown in Figure~\ref{app:fig:2x2_bursty_results}.

\begin{lstlisting}[
    float=htbp,
    caption={\lstinline{GenerateFlits} Process},
    label={app:cs:generate_flit_bursty},
    belowcaptionskip=0.5em,
    captionpos=b,
    language=Modest,
    style=ModestStyle
]
int[] burst_times = [0, 0, 0, 0];
int[] sleep_times = [0, 0, 0, 0];
const int BURST_MIN = 10;
const int BURST_MAX = 100;
const int SLEEP_MIN = 200;
const int SLEEP_MAX = 400;
process GenerateFlits(int id) {
 int(0..NOC_MAX_ID) destination;

 if (isBufferFull(noc[id].channels[LOCAL].buffer)) {
  generateFlits
 } else {
  // This custom process sends a flit in a bursty pattern
  if (burst_times[id] != 0) {
   generateFlits {=
    0: destination = DiscreteUniform(0, NOC_MAX_ID - 1),
    1: noc[id].channels[LOCAL].buffer = 
       enqueue(destination >= id ? 
         destination + 1 : 
         destination, noc[id].channels[LOCAL].buffer),
    2: burst_times[id] = burst_times[id] - 1
   =}
  } else if (sleep_times[id] != 0) {
   generateFlits {=
    sleep_times[id] = sleep_times[id] - 1
   =}
  } else {
   generateFlits {=
    burst_times[id] = DiscreteUniform(BURST_MIN, BURST_MAX),
    sleep_times[id] = DiscreteUniform(SLEEP_MIN, SLEEP_MAX)
   =}
}}}
\end{lstlisting}

\subsubsection{Prep Router Process}
\label{app:sec:processes:prep_router}

The channel datatype in Appendix Section \ref{app:sec:types:channel} has two Boolean flags that indicate if that channel is empty or if the channel is full. These are set in the \lstinline{PrepRouter} process (Code Segment~\ref{app:cs:prep_router_proc}) before flits are pushed to or popped from a buffer so that write-before-read conflicts do not occur.

\begin{lstlisting}[
    float=htbp,
    caption={\lstinline{PrepRouter} Process},
    label={app:cs:prep_router_proc},
    belowcaptionskip=0.5em,
    captionpos=b,
    language=Modest,
    style=ModestStyle
]
action prepRouter;
process PrepRouter(int id) {
 // Prep all the channels
 prepRouter {=
  // North Channel (0)
  noc[id].channels[NORTH].isEmpty
    = len(noc[id].channels[NORTH].buffer) == 0,
  noc[id].channels[NORTH].isFull
    = isBufferFull(noc[id].channels[NORTH].buffer),
  
  /* channels 1-4 */
 =}
}
\end{lstlisting}

\subsubsection{Advance Router Process}
\label{app:sec:processes:advance_router}

This process implements the routing algorithm described in Section \ref{sec:modular:flit_propogate}. Describing a template for this process is challenging, as it's very flexible in what routing algorithms can be implemented. The key steps that must be implemented in this process are (1) flits that have reached their destination must be removed and (2) flits destined for other routers must be moved toward their destination (as long as there is a channel they can move on).

This work implements an X-Y routing algorithm, as described in Section \ref{sec:routingAlg}. Additionally, this work uses a round-robin approach to arbiter traffic (Section \ref{sec:prelim:router_arb}). The \lstinline{priority_list} in the router datatype (Appendix Section \ref{app:sec:types:router}) keeps track of which buffers need to be serviced next, in order to ensure that all buffers are eventually serviced.

\begin{lstlisting}[
    float=htbp,
    caption={\lstinline{AdvanceRouter} Process},
    label={app:cs:advance_router_proc},
    belowcaptionskip=0.5em,
    captionpos=b,
    language=Modest,
    style=ModestStyle
]
process AdvanceRouter(int id) {
    AdvanceChannel(id, noc[id].priority_list[0]);
    AdvanceChannel(id, noc[id].priority_list[1]);
    AdvanceChannel(id, noc[id].priority_list[2]);
    AdvanceChannel(id, noc[id].priority_list[3]);
    AdvanceChannel(id, noc[id].priority_list[4])
}
\end{lstlisting}

Code Segment~\ref{app:cs:advance_router_proc} shows the the implementation of \lstinline{AdvanceRouter}. The advnace router process calls another sub-process \lstinline{AdvanceChannel} for each element in the priority list. The updating of the priority list occurs the \lstinline{UpdatePriority} (Appendix Section \ref{app:sec:processes:update_priority}). The specific \modest code of the \lstinline{AdvanceChannel} process are not given here, but detailed comments are available on \cite{noc_github} and algorithmic psuedocode is given in Algorithm~\ref{app:alg:advance_channel_xy} and~\ref{app:alg:advance_flits_xy}.

\begin{algorithm}
\caption{\lstinline{AdvanceChannel} Psuedocode}
\label{app:alg:advance_channel_xy}
\begin{algorithmic}[1]
\Require Current router ID $id$
\Require Current buffer from priority list $b$
\Ensure If possible, the front flit of buffer $b$ is either (1) removed if it's reached its destination, or (2) moved toward it's destination through a channel.

\Procedure{AdvanceChannel}{$id, b$}
    \If{$isNotConnected(b) \vee isEmpty(b)$}
        \State \Return
    \ElsIf{$reachedDestination(peekFront(b), id)$}
        \State $f \gets peekFront(b)$ \Comment{$f$ would be sent to the connecting hardware.}
        \State $b \gets dequeueFront(b)$ \Comment{$f$ gets dropped from $b$}
        \State \Return
    \Else \Comment{Flit is destined for another router}
        \State \Call{AdvanceFlits}{$id, b$} 
    \EndIf
\EndProcedure
\end{algorithmic}
\end{algorithm}

\begin{algorithm}
\caption{\lstinline{AdvanceFlits} Psuedocode with X-Y Routing}
\label{app:alg:advance_flits_xy}
\begin{algorithmic}[1]
\Require Current router ID $id$
\Require Current buffer from priority list $b$
\Require The front flit of $b$ ($peekFront(b)$) must not be destined for $id$
\Ensure If possible, a flit is moved $b$ into a neighboring routers input buffer according to the X-Y routing algorithm.

\Procedure{AdvanceFlits}{$id, b$}
    \State $\Delta x \gets columnOf(id) - columnOf(peekFront(b).id)$
    \State $\Delta y \gets rowOf(id) - rowOf(peekFront(b).id)$
    
    \If{$\Delta x = 0$}
        \If{$\Delta y > 0$}
            \State Send flit \textbf{\textit{north}}
        \Else
            \State Send flit \textbf{\textit{south}}
        \EndIf
    \ElsIf{$\Delta x > 0$}
        \State Send flit \textbf{\textit{east}}
    \Else
        \State Send flit \textbf{\textit{west}}
    \EndIf
\EndProcedure
\end{algorithmic}
\end{algorithm}

\subsubsection{Update Priority Process}
\label{app:sec:processes:update_priority}

This process updates the priority list member variable of the \lstinline{router} datatype and is used to arbitrate between conflicts between different buffers in a router. For example, if \rib{i}{South} and \rib{i}{North} each have a packet that is ready to send across \rib{i}{West} on the same clock cycle, the priority list will be used by the \lstinline{AdvanceRouter} (Appendix~\ref{app:sec:processes:advance_router}) to determine which packet is sent first.

A full discussion of the round robin semantics is given in \cite{boe2023probabilistic}.

\subsubsection{Clock Process}
\label{app:sec:processes:clock}

The \lstinline{clock} process is shown in Section~\ref{sec:modular:clock}, but the implementation is provided here as well in Code Segment~\ref{app:cs:Clock}.

\begin{lstlisting}[
    float=htbp,
    caption={\lstinline{Clock} Process},
    label={app:cs:Clock},
    belowcaptionskip=0.5em,
    captionpos=b,
    language=Modest,
    style=ModestStyle
]
process Clock() {
    nextClockCycle{= clk = (clk + 1) =};
    Clock()
}
\end{lstlisting}

\subsubsection{Update Noise Process}
\label{app:sec:processes:update_noise}

Code Segment~\ref{app:cs:update_noise_proc} shows how the counters \resist and \induct are updated. If the change in activity is above the threshold, then \induct is incremented and another inductive PSN event is considered to have happened. If the activity in the current cycle is above the threshold, then a resistive noise event is considered to have happened and \resist is incremented. Finally, the activity counters are updated to have the current activity now stored as the previous activity.

\begin{lstlisting}[
    float=htbp,
    caption={\lstinline{UpdateNoise} Process},
    label={app:cs:update_noise_proc},
    belowcaptionskip=0.5em,
    captionpos=b,
    language=Modest,
    style=ModestStyle
]
process UpdateGlobalNoiseTracking(int id) {
{= 
   // Update inductive noise
   0: inductiveNoise +=
    abs(noc[id].lastActivity - noc[id].thisActivity) >= ACTIVITY_THRESH
    ? 1 : 0,

   // Update resistive noise
   0: resistiveNoise += noc[id].thisActivity >= ACTIVITY_THRESH
    ? 1 : 0,

   // Update trackers for next round
   1: noc[id].lastActivity = noc[id].thisActivity,
   2: noc[id].thisActivity = 0
=}
}
\end{lstlisting}

\subsection{Generating Unique Model With Python}
\label{app:sec:python}

Due to some of the complexity of managing a modular NoC model in \modest, we created a Python script that allows for repeatable generation of unique NoC models. This script is not required to create new NoC models, and models can be updated by hand. However, it's simpler to automate this process to some extent, which is why we have provided the script. Code Segment~\ref{app:cs:noc_py_example} shows an example of generating a \nxn{2} NoC that can be used to characterize resistive noise with the \modestToolset.

\begin{lstlisting}[
    float=htbp,
    caption={NoC Python Library},
    label={app:cs:noc_py_example},
    belowcaptionskip=0.5em,
    captionpos=b
]
from noc import *
_2x2 = Noc(2, buffer_size = 4,
              activity_thresh = 3, ...)
with open("2x2.modest", "w") as f:
  f.write(_2x2.print(PropertyType.RESISTIVE))
\end{lstlisting}

The \lstinline{Noc} class is available in \cite{noc_github}.

\subsection{Adding a Unique Flit Generation Algorithm to the Model}

Additionally, functionality is in place for unique flit generation algorithms to be added to models generated using the NoC class as shown in Code Segment~\ref{app:cs:python_generate_flits_bursty}. The unique flit generation algorithm must meet the requirements detailed in Section~\ref{sec:modular:flit_gen} and Appendix~\ref{app:sec:processes:generate_flits}, and should be encoded in plain text as a string in Python. Then to generate the model using the custom flit generation process, simply pass the string in to the Noc class \lstinline{print} method to substitute the custom process in.

\begin{lstlisting}[
    float=htbp,
    caption={Using the Python Library to Generate Custom Flits},
    label={app:cs:python_generate_flits_bursty},
    belowcaptionskip=0.5em,
    captionpos=b
]
from noc import *
    
custom_generate_flits: str = """
int[] burst_times = [0, 0, 0, 0];
int[] sleep_times = [0, 0, 0, 0];
const int BURST_MIN = 10;
const int BURST_MAX = 100;
const int SLEEP_MIN = 200;
const int SLEEP_MAX = 400;
process GenerateFlits(int id) {
process GenerateFlits(int id) {
 int(0..NOC_MAX_ID) destination;

 if (isBufferFull(noc[id].channels[LOCAL].buffer)) {
  generateFlits
 } else {
  // This custom process sends a flit in a bursty pattern
  if (burst_times[id] != 0) {
   generateFlits {=
    0: destination = DiscreteUniform(0, NOC_MAX_ID - 1),
    1: noc[id].channels[LOCAL].buffer = 
       enqueue(destination >= id ? 
         destination + 1 : 
         destination, noc[id].channels[LOCAL].buffer),
    2: burst_times[id] = burst_times[id] - 1
   =}
  } else if (sleep_times[id] != 0) {
   generateFlits {=
    sleep_times[id] = sleep_times[id] - 1
   =}
  } else {
   generateFlits {=
    burst_times[id] = DiscreteUniform(BURST_MIN, BURST_MAX),
    sleep_times[id] = DiscreteUniform(SLEEP_MIN, SLEEP_MAX)
   =}
  }
 }
}
"""

_2x2 = Noc(2);
_2x2_as_str = _2x2.print(PropertyType.RESISTIVE,
                         generate_flits=custom_generate_flits)
\end{lstlisting}

\end{document}